\documentstyle[12pt,epsfig]{article}

\begin{document}

\begin{center}
{\large IN QUEST OF NEUTRINO MASSES AT ${\cal{O}}$(eV) SCALE} \\
\vspace{1 cm}
 M. CZAKON \\
\vspace{.3 cm}
Institut f\"ur Theoretische Physik, Universit\"at Karlsruhe, \\
D-76128 Karlsruhe, Germany \\
\vspace{.5 cm}
J. GLUZA J. STUDNIK AND M. ZRA$\L$EK \\
\vspace{.3 cm}
Department of Field Theory and Particle Physics,
\\
                Institute of Physics, University of Silesia,
\\
                Uniwersytecka 4, PL-40-007 Katowice, Poland
\end{center}
%\maketitle

\begin{abstract}

Neutrino oscillation and  tritium beta decay experiments
taken simultaneously into account are able to access the  so far
imperceptible absolute neutrino masses
at the electronvolt level. The neutrino mass spectrum derived in this
way is independent of the nature of neutrinos (Dirac or Majorana).
Furthermore, the  lack of neutrinoless double beta decay gives additional
constraints on the Majorana neutrino mass spectrum.
A case of three neutrinos is examined.
Influence of different solutions to the solar neutrino deficit problem 
on the results is discussed. Apart from the present situation,
four qualitatively distinct experimental situations which are possible 
in the future are investigated: when the  two decay experiments give 
only upper bounds on effective neutrino masses, when either one of them 
gives a positive result, and when both give positive results.
The discussion is carried out by taking into account
the present experimental errors of relevant neutrino parameters
as well as their much more precise expected estimations
(e.g. by $\nu$ factories). It is shown in which cases the upgraded decay 
experiments simultaneously with neutrino oscillation data
may be able to fix the absolute scale of the neutrino mass spectrum,
answer the question of the neutrino nature and put some light
on CP phases in the lepton sector.
\end{abstract}

\newpage
%\vspace{.5cm}

\section{Introduction}
%\subsection{Basic facts}
The problem of the neutrino mass spectrum and its nature is the most
important issue in the lepton part of
the Standard Model. What new information can we obtain from the last
experimental results, and what are the
future perspectives? Three kinds of experiments play a fundamental role
in answering this question. Two are
traditional and known for years: beta decay and neutrinoless double beta
decay $(\beta\beta)_{0\nu}$ of
nuclei. Already Fermi \cite{fermi} in 1934 and Furry \cite{furry} in
1939 realized
that both processes are important for  the
neutrino mass and nature. The third type
constitute the neutrino oscillation experiments.
These are responsible for anomalies observed in
solar \cite{davis}, atmospheric \cite{fukuda} and LSND
\cite{athanassopoulos} experiments.
Though trials of alternative explanations of the observations
 exist \cite{pakvasa},
they require much more sophisticated
assumptions (as for example the breaking of the equivalence principle,
breaking of
the special theory of relativity, the
neutrino decay with life-time much below expectations or huge neutrino
magnetic moments) and give much
poorer fits to the data \cite{lipari}\footnote{There are also some
astrophysical and cosmological arguments which put some light on
the neutrino
masses. One of them comes from the analysis relating the
cosmic microwave background temperature fluctuations to
the present large scale structure formation. It depends strongly on the
accepted cosmological model of the Universe and for three light
neutrinos gives $m_{\nu}\leq 1.8(0.6)$ eV for any value of the
cosmological density $\Omega_m$ ($\Omega_m=0.3$)
 \cite{cmb}. Another bound comes from the observation of ultrahigh
cosmic rays. The so-callled Z-burst model \cite{zburst}
gives $m_{\nu} \in (0.1 \div 1)$ eV.
Though the above numbers are very impressive (and better than the present
tritium $\beta $ decay bound), they depend on
additional assumptions connected to the interpretation of astrophysical
data
and we will not include them in the present analysis (see \cite{weiler}
and \cite{apbgen} for a discussion which includes the cosmological
data).}.

Solar, atmospheric  and LSND
experiments probe the 
neutrino oscillation hypothesis with three disconnected ranges of
$\delta m^2$ parameters ($\delta m^2\simeq  10^{-10} \div 10^{-5}$ $eV^2$ 
for solar neutrinos,   $\delta m^2\simeq  10^{-3} \div 10^{-2}$ $eV^2$ 
for atmospheric neutrinos and $\delta m^2\simeq  0.1 \div 10$ $eV^2$ 
for the LSND experiment). The situation seems
to be clear and in favor of the neutrino oscillation hypothesis  for the
atmospheric neutrino data analysis (agreement among different
experiments). Also, the  solar neutrino deficit is quite well explained by
neutrino oscillations, but at present no unique  solution
for the oscillation parameters exists. As far as LSND is concerned, 
the situation is
currently not clear at all \cite{notcl}. The LSND results, if
confirmed, would imply a fourth, sterile neutrino. Here we consider
mass scenarios with three neutrinos only.

\begin{figure}[t]
\begin{center}

\epsfig{figure=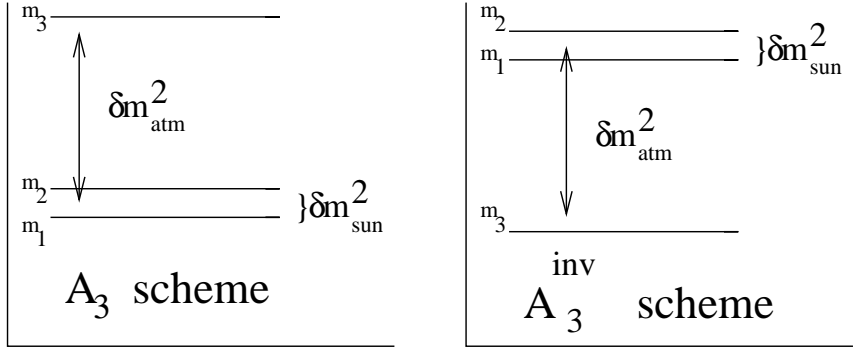, scale=1.2}

\caption{Two possible mass spectra which can describe the oscillation
data. Scheme $A_3$, normal mass hierarchy,
has a small gap between $m_1$ and $m_2$ to explain the oscillation of
solar neutrinos and a larger gap
for the atmospheric neutrinos ( $\delta m^2_{sol}=\delta m^2_{21}<<
\delta m^2_{32} \simeq  \delta m^2_{atm}$;
$m_1<m_2<<m_3$). In the inverse mass hierarchy scheme $A_3^{inv}$,
$-\delta m^{2}_{31}=\delta m^{2}_{atm}>>\delta m^{2}_{solar} \approx
\delta m^{2}_{21}$
The mass of the lightest neutrino $(m_{\nu})_{min}=m_{1}$ in the $A_{3}$
schemes  and $(m_{\nu})_{min}=m_{3}$ in the $A^{inv}_{3}$ scheme.}
\label{spectra}
\end{center}
\end{figure}

As there are definitely two scales of $\delta m^2$, $\delta m^2_{atm}
\gg
\delta m^2_{sol}$, two possible neutrino mass spectra must
be considered (Fig.~\ref{spectra}). The first, known
as normal mass hierarchy ($A_3$) where $\delta m^2_{sol} = \delta
m^2_{21} \ll \delta m^2_{32} \approx \delta
m^2_{atm}$ and the second, inverse mass hierarchy spectrum ($A^{inv}_3$)
with $\delta m^2_{sol} = \delta
m^2_{21} \ll -\delta m^2_{31} = \delta m^2_{atm}$,  $\delta
m_{ij}^{2}=m_{i}^{2}-m_{j}^{2}$. Both schemes are not distinguishable by
present
experiments. There is hope that next long base line experiments (e.g.
MINOS, ICARUS) and/or
neutrino factories will do that\footnote{According to a recent
analysis of the neutrino spectrum from the SN1987A \cite{sn1987}, the
$A^{inv}_3$ scheme is disfavored for $|U_{e3}|^2>0.001$.}.
Such schemes are the basic ones. As the neutrino mass spectrum is
determined
by the mass of the lightest
neutrino $(m_{\nu})_{min}$,
other possible neutrino mass schemes
known in the literature as ``quasi degenerate'' ,
``partial mass hierarchy''or ``partial inverted mass hierarchy''
\cite{many,last} are considered automatically
in the paper ($(m_{\nu})_{min}$ in the range from  zero up to around 2.2
eV
is taken into account).
%as a parameter all other schemes as ``quasi degenerate'', ``partial
%mass hierarchy'' \cite{Petkov} are considered automatically.

The oscillation experiments are able to find differences of mass squares
$\delta m^2$ (not the absolute masses separately) and absolute values of
some of the mixing matrix elements  $|U_{ei}|$ (presently no information on
CP phases is available).

%Present values of the allowed ranges of $\delta m^2$ and  $\sin^2{2
%\theta}$ (at 95 \% c.l.) for atmospheric and solar oscillation
%parameters are given in  Table~1.

Different combinations of masses and  $U_{ei}$'s are measured in tritium
$\beta$ and $(\beta\beta)_{0\nu}$ decays. Taking these data
together we can probe the absolute neutrino masses. Such an analysis has
been partially done in different contexts in
\cite{apbgen,many,last,16a,plb1}.

Here, our main motivation is to answer the following questions:
when, how precisely and under what circumstances the absolute neutrino
masses can be determined.
As can be expected, the answer depends crucially on the mass of the
lightest neutrino  $(m_{\nu})_{min}$.
For  $(m_{\nu})_{min}$ above approximately 0.3 eV
(the exact value which is discussed later on,
depends on the precision of the neutrino oscillation parameters'
determination)
%($\sin^2{2 \Theta_{sol,atm}}$, $|U_{e3}|^2$, $\delta m^2_{sol,atm}$))
we can expect that the upgraded
tritium $\beta$ decay experiments together with the oscillation
data will
be able to determine
the absolute  neutrino masses $m_{i}$ independently of the neutrino
character.
If $ (m_{\nu})_{min} \leq  0.3$ eV, the $(\beta\beta)_{0\nu}$ decay
gives some chance to determine  $(m_{\nu})_{min}$.
We discuss the conditions required for this to happen. 
Two future scenarios are considered.
In the first case neutrinos are Majorana particles and the
effective Majorana mass is determined by the $(\beta\beta)_{0\nu}$
experiment.
In the second case  the nature of neutrinos is not known
and we will still have only a bound on $\left< m_{\nu} \right>$.
In this case there are circumstances when the combined results
from $(\beta\beta)_{0\nu}$,  tritium $\beta$ decay
and oscillation experiments are able to exclude the Majorana nature of the
neutrino.
In our numerical analysis, special
attention is paid to the influence of present and future
experimental errors on the absolute neutrino mass
determination.
It is shown that the expected improvements from incoming
$\nu$ factories will provide additional severe constraints on the neutrino
masses.
%These considerations constitute the main results of the paper.
%It is shown that expected precision to which relevant neutrino
%parameters' will be determined cause  of neutrino
%parameters in future we

The paper is organized as follows.
In the next Section, the experimental status
of neutrino oscillation searches, $\beta$ tritium and
$(\beta\beta)_{0\nu}$ decays
is shortly reviewed. Expected improvements of the precision
of parameters' determination are listed.
In Section 3 basic analytical formulae
which are used in the neutrino absolute mass search are presented.
%Next Section gives
%experimental data on neutrino masses and mixings needed
%for further discussion are given.
Section 4
includes a discussion of numerical results.
Four possible future scenarios  mentioned in the Abstract and
their consequences for the determination of the
neutrino mass spectrum are analyzed.
The paper ends  with conclusions.

\section{Main neutrino data and their present
and future experimental precision}

%\subsection{Oscillation experiments}

A global analysis of the solar, atmospheric and reactor neutrino
data determines five parameters: three mixing angles
($\Theta_{12},\Theta_{13},
\Theta_{23}$ ($0 < \Theta_{ij} < \frac{\pi}{2}$)
and two mass square differences  ($\pm \delta m_{31}^{2}= \delta
m^2_{atm}>0$,
$\pm \delta m_{21}^2=\delta m^2_{sol}>0$).

For the solar neutrino problem several analyses have been
carried out so far allowing mixings among 2,3 or 4 neutrinos
\cite{2n,3n,4n}. The results differ slightly,
nevertheless, in the 3$\nu$ scenario
four solutions for the solar neutrino deficit
are still acceptable at the 95\% c.l. \cite{3n}.
The first one, is the small mixing angle solution (SMA MSW) with
$\sin^2{2\Theta} \simeq 0.001 \div 0.01$. Three remaining solutions
(LMA MSW, LOW MSW and QVO) includes
large mixing angles, namely  $\sin^2{2
\Theta} \geq 0.55$.
In these cases a maximal mixing  $\sin^2{2
\Theta}=1$ is still acceptable.
%As we would like to consider deviations
%from maximal mixing, a parameter $\epsilon=1-\sin^2{2\Theta}$ is
%introduced. Our analysis depends very weakly on  $\delta m^2_{atm}$
%which
%are different for LMA MSW, LOW MSW and VO solutions. In such
%circumstances all three large mixing angle solutions can be considered
%%as a one case. That is why we separate our discussion to two cases:
%SMA
%MSW and LMA($\epsilon$) MSW.
The present situation  and future expectations are summarized in
Table~1.
The matter enhanced solution of the solar neutrino problem is accepted
for $\delta m _{21}^{2}>0$ only.
The sign of $\delta m _{31}^{2}$ cannot yet be determined, so two
schemes are considered (Fig.~\ref{spectra}). Incoming long baseline
experiments and especially neutrino factories
should be able to distinguish between these two schemes.

The mixing angles given in Table~1 enter the
effective neutrino mass formulae, which can be written as,

\begin{equation}
\label{mbeta}
m_{\beta} \equiv \left[ \sum^3_{i=1} |U_{ei}|^2m^2_i\right]^{1/2}
\end{equation}
for the tritium $\beta$ decay, and
% \footnote{
%Another definition of the  effective $m_{\beta}$ parameter has been
%recently proposed
%in \cite{mbeta2u}
%($m_{\beta}=\sum\limits_i| U_{ei}|^2 m_i$). }

% which is
% independently of the relation between neutrino masses $m_{i}$ and the
%energy interval $\Delta E$
%end of electron spectrum, resolved by the detector parameter given in
%form (~\ref{mbeta}) is better. Only one assumption which we read is%
%$\sqrt{\delta m_{ij}^{2}} \ll \Delta E$ and

\begin{equation}
\left< m_{\nu} \right> =|\sum\limits_i U_{ei}^2 m_i|
\label{bb}
\end{equation}
for the $(\beta\beta)_{0\nu}$ decay.

In both schemes $U_{ei}$'s are given by
\begin{equation}
U_{e1}=\cos{\Theta_{12}}\cos{\Theta_{13}}, \;
U_{e2}=\sin{\Theta_{12}}\cos{\Theta_{13}}, \;
|U_{e3}|=\sin{\Theta_{13}}.
\label{trzy}
\end{equation}

The experimental data at the end of the Curie plot in the tritium $\beta$
decay provides the upper
limit  on the effective electron neutrino mass  $m_\beta$.

The present best limit is given by the Mainz collaboration \cite{weinheimer}
\begin{equation}
m_\beta  <  \kappa'= 2.2 \; \mbox{\rm eV}.
\label{trit}
\end{equation}

A second collaboration from Troitsk gives similar results \cite{lobashev}

\begin{equation}
m_\beta  <   2.5 \; \mbox{\rm eV}.
\label{troitsk}
\end{equation}

The groups from Mainz, Troitsk, Karlsruhe and Fulda have presented a
project \cite{katr}
for a new experiment (KATRIN), which should  improve the existing limit
by a factor of ten, so within
6-7 years $m_\beta$ should reach $m_\beta \sim $ 0.3 eV.

The effective neutrino mass $\left< m_{\nu} \right>$
in Eq.~\ref{bb} is extracted from
the decay half life time of even-even nuclei
\cite{half}
\begin{equation}
{\left[ T^{1/2} (\beta\beta)_{0\nu} \right]}^{-1}=|M_{nucl}|^2 \times
\left( \mbox{\rm Phase space integral} \right)
\times \frac{ \left<m_{\nu} \right>^2}{m_e^2}+...
\end{equation}
Dots represent the other, different from direct light Majorana neutrino exchange
mechanisms
which can contribute to $ (\beta\beta)_{0\nu}$ decay (e.g.
mechanisms with   heavy neutrinos  or supersymmetric particles \cite{other}).
The identification of various mechanisms responsible
for the neutrinoless double beta decay, as well as the precise calculation
of the nuclear matrix elements is a very difficult task, e.g.
the nuclear matrix element $|M_{nucl}|^2$ has been
calculated by several groups and the results differ among them
roughly by a factor of 3 \cite{half,other,diff}.

The present limit on the effective light neutrino mass is
\cite{baudis}

 \begin{equation}
\left<m_{\nu}\right> < \kappa = 0.2\; \mbox{\rm eV }.
\label{pres}
\end{equation}

Several new experiments are considered which will be able to further increase
the  sensitivity of the $\left<m_{\nu}\right>$
measurement
\cite{nemo3,ouore,moon} though
the best limit is planned to be obtained by the GENIUS experiment.
In its first stage of running,
GENIUS with 1 ton of $^{76} Ge$ should be able to reach a sensitivity of $
\left<m_{\nu}\right> \sim 0.02 \ eV$,
later with 10 tons of $^{76} Ge$,  a sensitivity of the order of
$\left<m_{\nu} \right> \sim 0.006 \ eV$ will be available
\cite{exo}.

\section{Dirac and Majorana neutrino masses in  the $A_3$ and  $A_3^{inv}$
schemes: analytical formulae}

Here we summarize the key expressions, used to determine the absolute neutrino
masses. It is known that the electron energy distribution in the $\beta$
decay of nuclei and flavour oscillations do not distinguish between
Dirac and Majorana neutrinos.

{\bf I.}
Oscillation experiments\\

Since in both neutrino mass schemes
\begin{equation}
(m_{\nu})_{max}^2 = (m^2_{\nu})_{min}+\delta m^2_{solar}+\delta
m^2_{atm},
\label{m3}
\end{equation}
the oscillation experiments alone give  ~\cite{barger}

\begin{equation}
(m_{\nu})_{max} \geq \sqrt{\delta m^{2}_{solar}+\delta m^{2}_{atm}},
\label{m}
\end{equation}
and
\begin{equation}
|m_i-m_j| \leq \sqrt{\delta m^2_{solar}+\delta m^2_{atm}}.
\label{ij}
\end{equation}

{\bf II.}
Tritium $\beta$ decay\\

From the effective neutrino mass formula Eq.~\ref{mbeta} we can find
a double inequality

\begin{equation}
(m_{\nu})_{min} \leq m_\beta \leq(m_{\nu})_{max}.
\end{equation}
Currently, we have  only a bound on  $m_{\beta}$ Eq.~\ref{trit}, that
gives a limit on the absolute neutrino mass
\begin{equation}
\label{mnumin}
 (m_{\nu})_{min} \leq \kappa'
\end{equation}
%$m_{3}$ remains unfortunately unlimited from above.
without any limits on $(m_{\nu})_{max}$.\\

{\bf III.}
The tritium $ m _{\beta} $ decay together with neutrino oscillation data\\

From  Eq.~\ref{mbeta} we can find the relations
\begin{equation}
\label{omega}
m_{ \beta}^2 = (m_{\nu})_{min}^{2}+ \Omega_{scheme},
\end{equation}
and

\begin{equation}
(m_{\nu})_{max}^{2} = m^2_\beta+\Lambda_{scheme},
\label{lambda}
\end{equation}
where $\Omega$ and $\Lambda$ are scheme dependent quantities and in both
schemes $A_{3}$ and $A_{3}^{inv}$ are given by

\begin{eqnarray}
\Omega(A_3) &=& (1-|U_{e1}|^2)\delta m^2_{solar} + |U_{e3}|^2 \delta
m^2_{atm},\label{pie} \\
\Lambda(A_3)& =& |U_{e1}|^2 \delta m^2_{solar}+(1-|U_{e3}|^2) \delta
m^2_{atm}, \\
\Omega(A_3^{inv}) &=& (1-|U_{e3}|^2)\delta m^2_{atm} + |U_{e1}|^2 \delta
m^2_{solar}, \label{omegainv} \\
\Lambda(A_3^{inv})& =& (1-|U_{e1}|^2)\delta m^2_{solar} + |U_{e3}|^2
\delta m^2_{atm}. \label{ost}
\end{eqnarray}
From Eqs.~\ref{trit},\ref{m3},\ref{omega},\ref{lambda}
better limits on $(m_{\nu})_{min}$ and $(m_{\nu})_{max}$ follow

\begin{eqnarray}
 (m_{\nu})_{min} & \leq & \sqrt{(\kappa ')^2-\Omega^{min}_{scheme}},
\\
\label{big}
\sqrt{\delta m^2_{solar}+\delta m^2_{atm}} & \leq & (m_{\nu})_{max}
\leq \sqrt{(\kappa ')^2+\Lambda^{max}_{scheme}},
\end{eqnarray}
where $\Omega^{min}_{scheme}$ and $\Lambda^{max}_{scheme}$
are the allowed minimal and maximal values given by
Eqs.~(\ref{pie})-(\ref{ost}).\\
We can see that the knowledge of $m_{\beta}$ together with the
oscillation parameters gives a simple
way to determine the absolute neutrino masses. If the neutrino happens to be
a Dirac particle, then this  will be the only way to determine its mass.

If neutrinos are Majorana particles the bound on $ \left< m_{\nu}
\right>$ applies and additional  constraints follow.\\

{\bf IV.}
Neutrinoless double beta decay\\

For three neutrinos in the  $A_{3}$ and $A_{3}^{inv}$ schemes
 we have

\begin{eqnarray}
\label{basic}
\langle m_\nu \rangle _{A_{3}} &=& \left| |U_{e1}|^2
(m_\nu)_{min}^{2}+|U_{e2}|^2
e^{2i\phi_2} \sqrt{(m_{\nu})_{min}^2+\delta m^2_{sol}} \right. \nonumber
\\
&+& \left. |U_{e3}|^2 e^{2i\phi_3} \sqrt{(m_{\nu})_{min}^2+\delta
m^2_{sol}+\delta m^2_{atm}} \right| ,
\end{eqnarray}
and
\begin{eqnarray}
\label{basic1}
\langle m_\nu \rangle _{A_{3}^{inv}} &=& \left| |U_{e1}|^2
\sqrt{(m_{\nu})_{min}^2+\delta m^2_{atm}}+|U_{e2}|^2
e^{2i\phi_2} \times  \right. \nonumber \\
& \times &\sqrt{(m_{\nu})_{min}^2+\delta m^2_{atm}+\delta m^2_{sol}}
\left.+ |U_{e3}|^2 e^{2i\phi_3} (m_{\nu})_{min}^2\right| .
\end{eqnarray}
The three
parameters  used above, $(m_{\nu})_{min}$ and two Majorana
$CP$ violating phases $\phi_1$ and $\phi_2$ are unknown. We
are not able to
predict the value of $ \left<m_{\nu} \right> $ as a
function of $(m_{\nu})_{min}$ but a range
%(using appropriate phases)
\begin{eqnarray}
\label{minmax}
(\left< m_{\nu} \right>_{min}, \left< m_{\nu} \right> _{max}) ,
\end{eqnarray}
can be obtained \cite{apbgen}.

At present only the upper bound on $ \left< m_{\nu} \right> $
(Eq.~\ref{pres}) is known.
This result allows us to estimate the minimal mass of the lightest
neutrino  $(m_{\nu})_{min}$.
In future, if the $(\beta \beta)_{0 \nu}$ experiment gives a positive result
and a value $ \left< m_{\nu} \right>= \kappa \pm \Delta \kappa$
is found, the problem of the neutrino mass determination depends on the 
relation between the range given in Eq.~(\ref{minmax}) and the value of
$\kappa$. Such a possibility will be discussed later.

\section{Absolute neutrino masses: numerical results}

\subsection{Dirac or Majorana case}

From oscillation experiments  we can only state that the mass of the heaviest
neutrino must be larger than 0.04 eV (Eq.~\ref{m}, Table~1)
\begin{eqnarray}
\label{.04}
(m_{\nu})_{max} \geq 0.04 \  eV,
\end{eqnarray}
and the difference between any two neutrino masses is smaller than 0.08
eV
\begin{eqnarray}
\label{.08}
|m_{i}-m_{j}| < 0.08 \ eV.
\end{eqnarray}

These results depend on the precision of $ \delta m^{2}_{atm}$
(Eqs.~\ref{m},\ref{ij}).
It means that a
future improvement in the determination of $ \delta m^{2}_{atm}$ (up to
$ 1 \%$, see Table ~1)
will result in a substantial improvement of these bounds.

From the tritium $\beta$-decay (Eqs.~\ref{trit},\ref{mnumin}) we can find
that the mass of the lightest neutrino must be smaller than 2.2 eV
\begin{eqnarray}
\label{2.2}
(m_{\nu})_{min} < 2.2 \ eV  ,
\end{eqnarray}
which together with the bound Eq.~\ref{.08} gives limits on the masses of each
neutrino separately
\begin{eqnarray}
\label{mi}
m_{i} \leq 2.2 \  eV, \ \ \ \ \ \  i=1,2,3.
\end{eqnarray}
Eqs.~\ref{.04},~\ref{.08},\ref{mi} establishes the present knowledge of the 
neutrino masses independently of their nature.
For Dirac neutrinos there is no better sources of information.
In future the  $_{1}^{3}H$ decay supplemented by the oscillation data
will be able to
 reconstruct the Dirac or Majorana neutrino mass spectrum up to small
values of $(m_{\nu})_{min}$. This can be done quite precisely. 
From Eq.~\ref{omega} it follows that the relative error of
$(m_{\nu})_{min}$ is given by
\begin{eqnarray}
\label{stos}
\frac{\Delta
(m_{\nu})_{min}}{(m_{\nu})_{min}}&=&\frac{m_{\beta}}{(m_{\nu})_{min}^{2}}
\Delta m_{\beta}+\frac{1}{2(m_{\nu})_{min}^{2}} \Delta
(\Omega_{scheme}) 
\end{eqnarray}
The part of $\Delta (m_{\nu})_{min}$ which comes from the uncertainties
of neutrino oscillation parameters is very small
\begin{eqnarray}
\label{deltaomega}
\Delta (\Omega_{A_{3}})=3.4 \times 10^{-4}, \\ \nonumber
\Delta (\Omega_{A_{3}^{inv}})=29.4  \times 10^{-4},
\end{eqnarray}
and in the range of the KATRIN experiment ($(m_{\nu})_{min} \sim  0.3$ eV
\cite{katr}) the error becomes negligible
\begin{eqnarray}
\label{pr}
\frac{ \Delta (\Omega_{A_3})}{2(m_{\nu})_{min}^{2}} & \approx & 0.2 \%,
\\
\frac{ \Delta (\Omega_{A_3^{inv}})}{2(m_{\nu})_{min}^{2}} & \approx &
1.6 \%.  \\
\end{eqnarray}
%\left(\frac{\Delta (m_{\nu})_{min}}{(m_{\nu})_{min}} \right)_{oscill.}
%\approx 0.2 \%  \ \ \ (A_{3}) \\ \nonumber
%\ \ \ \  \ \ \ \ \ \ \ \ \ \ \ \ \ \ \ \ \ \ \ \ \  \approx 1.6 \%  \ \
%\ ( A_{3}^{inv})

The errors increase with decreasing $(m_{\nu})_{min}$,
e.g. for $(m_{\nu})_{min}=0.13$ eV the error is 1 \%
($A_3$ scheme).
%To get in $A_{3}$ case an error
%In $A_{3}$ case this error becomes miningfull only for smaller neutrino
%masses , for example
%\begin{eqnarray}
%\label{ap}
%\left(\frac{\Delta (m_{\nu})_{min}}{(m_{\nu})_{min}}\right)_{oscill.}
%\approx 1\%  \ \ \ for \ \ \ (m_{\nu})_{min}=0.13 \ eV
%\end{eqnarray}
Future improvements in the determination of neutrino oscillation parameters
will decrease this error substantially, e.g., using the
estimations from the last column of Table~1
\begin{eqnarray}
\label{ap}
\frac{ \Delta (\Omega_{A_3})}{2(m_{\nu})_{min}^{2}} & \approx & 1\%  \ \
\ \mbox{\rm for} \ \ \ (m_{\nu})_{min}=0.02 \ eV.
\end{eqnarray}

As we can see the main error comes from $\Delta m_\beta$ which will  be
under control for
$m_{\beta} \geq 0.3$ eV. Since in this case the error connected to
uncertainties of the oscillation parameters
is below 1 \%,
the tritium $\beta$ decay together with the oscillation
experiments would be the
ideal place for the neutrino mass spectrum reconstruction as long as
$(m_{\nu})_{min}>0.3$ eV.
If neutrinos are Dirac particles and their masses are below this scale,
then the absolute neutrino mass
determination seems to be out of  reach, unless some new methods of
direct neutrino mass measurements are developed.

\subsection{Majorana case}

Currently, the bound on the effective Majorana mass $ \left< m_{\nu}
\right>$ is one order of magnitude better  than on
$m_{\beta}$ (compare Eq.~\ref{trit} and Eq.~\ref{pres}). Moreover,
there are really impressive plans to get
$ \left< m_{\nu} \right> \approx 0.006 \  eV$ in
$(\beta \beta)_{0 \nu}$  experiments.
Will they be able to get down with a sensitivity of $(m_{\nu})_{min}$ to
the meV scale?
The situation seems to be very promising, however  $ \left< m_{\nu}
\right> $ depends on the
 Majorana phases which can lead to large  cancellations.
For this reason, the range of  possible $ \left< m_{\nu} \right>$ values
(Eq.~\ref{minmax}) can be very wide.
%\begin{eqnarray}
%\label{przedz}
%(<m_{\nu}>_{min},<m_{\nu}>_{max})
%\end{eqnarray}
This range depends also very crucially on the $U_{ei}$
mixing matrix elements which are not known with a satisfactory
precision (see Table ~1, Eq.~\ref{trzy}).
%Solar neutrino experiments, which determine $U_{e1}$ and $U_{e2}$, have
%still three broad regions
%of possible mixing angles $\theta_{12}$ (Table ~1).
The reactor experiments which determine $U_{e3}$ are not enough
precise, namely \cite{chooz}
\begin{eqnarray}
\label{mm}
|U_{e3}|^{2} \leq 0.04 .
\end{eqnarray}

The maximum of  $ \left< m_{\nu} \right> $ is stable and depends
mostly on $\theta_{13}$

\begin{eqnarray}
\label{4lin}
\left< m_{\nu} \right>_{max}&=&(cos^{2} \theta_{12} m_{1}+sin^{2}
\theta_{12} m_{2}) cos^{2} \theta_{13}+m_{3} sin^{2} \theta_{13},
\end{eqnarray}
so for  various regions of masses there is
\begin{eqnarray}
 A_3: \;\;\;\;\;\;
 \left< m_{\nu} \right>_{max } &\approx&  m_{3} sin^{2} \theta_{13}, \ \
\ \ \ \ \ \ \ \ \ \ \ \ \ \ \ \ \ \ \  \ \ m_{1}<<m_{2}<<m_{3} \nonumber
\\
 \ \  &\approx& m_{1} cos^{2} \theta_{13}+m_{3} sin^{2} \theta_{13}, \ \
\ \  \ \ m_{1} \approx m_{2}<<m_{3}  \nonumber\\
 \ \ &\approx& m_{1}, \ \ \ \ \ \ \ \ \ \ \ \ \ \ \ \ \ \ \ \ \ \ \ \ \
\ \ \ \ \  \ \ m_{1} \approx m_{2} \approx m_{3}  \\
\label{1lin}
A_3^{inv}:  \;\;\;\;\;\;
\left< m_{\nu} \right>_{max} &\approx& cos^{2} \theta_{13} m_{1},  \ \ \
\ \ \ \ \ \ \ \ \ \ \ \ \ \ \ \  \ \ m_{1} \approx m_{2} \gg m_{3}
\nonumber\\
&\approx& m_{1}, \ \ \ \ \ \ \  \ \ \ \ \ \ \ \ \ \ \ \ \ \ \ \ \ \ \ \
\  \ \ m_{1} \approx
 m_{2} \approx m_{3}.  \nonumber
\end{eqnarray}

The formula which gives the minimal value of $ \left< m_{\nu} \right>$
is much more complicated and strongly depends on the  solar mixing angle
$\theta_{12}$. If $\theta_{12} \approx \frac{\pi}{4}$, cancellations
among  all three terms in $ \left<m_{\nu} \right>$
are possible and $ \left< m_{\nu} \right>_{min}$ can be negligible
small.
If $\theta_{12} \neq \frac{\pi}{4}$, one of two terms $|U_{e1}|^{2}
m_{1}$,
$|U_{e2}|^{2} m_{2}$  dominates the cancellation is not complete, 
$\left< m_{\nu} \right>_{min}>0$.
To see it let us take a large value of $(m_{\nu})_{\min}$
($(m_{\nu})_{\min} \gg \delta m_{atm}^{2}$), then in both schemes
\begin{eqnarray}
\label{sc}
\left< m_{\nu} \right>_{min}& \simeq &(m_{\nu})_{min} (cos^{2}
\theta_{13}|cos^{2} \theta_{12}-sin^{2} \theta _{12}|-sin^{2} \theta
_{13})= \\ \nonumber
&=&  (m_{\nu})_{min}(\epsilon cos^{2} \theta_{13}-sin^{2} \theta _{13}),
\end{eqnarray}
where the parameter $ \epsilon$\footnote{The formula (Eq.~\ref{sc}) is
valid only for $ \epsilon > \tan^{2} \theta _{13}$.
For smaller values of $ \epsilon$, $ \left<m_{ \nu} \right>_{min}=0$
\cite{apbgen}.}  is introduced
\begin{eqnarray}
\label{epsilon2}
\epsilon =|cos^{2} \theta_{12}-sin^{2} \theta_{12}|=
\sqrt{1-sin^{2}2 \Theta_{12}}.
\end{eqnarray}

This new parameter  measures the 
deviation of the $\Theta_{12}$ angle from
its maximal value ($\theta_{12}= \frac{\pi}{4}$) and is quite suitable
for our discussion.

\begin{figure}[t]
\begin{center}
\epsfig{figure=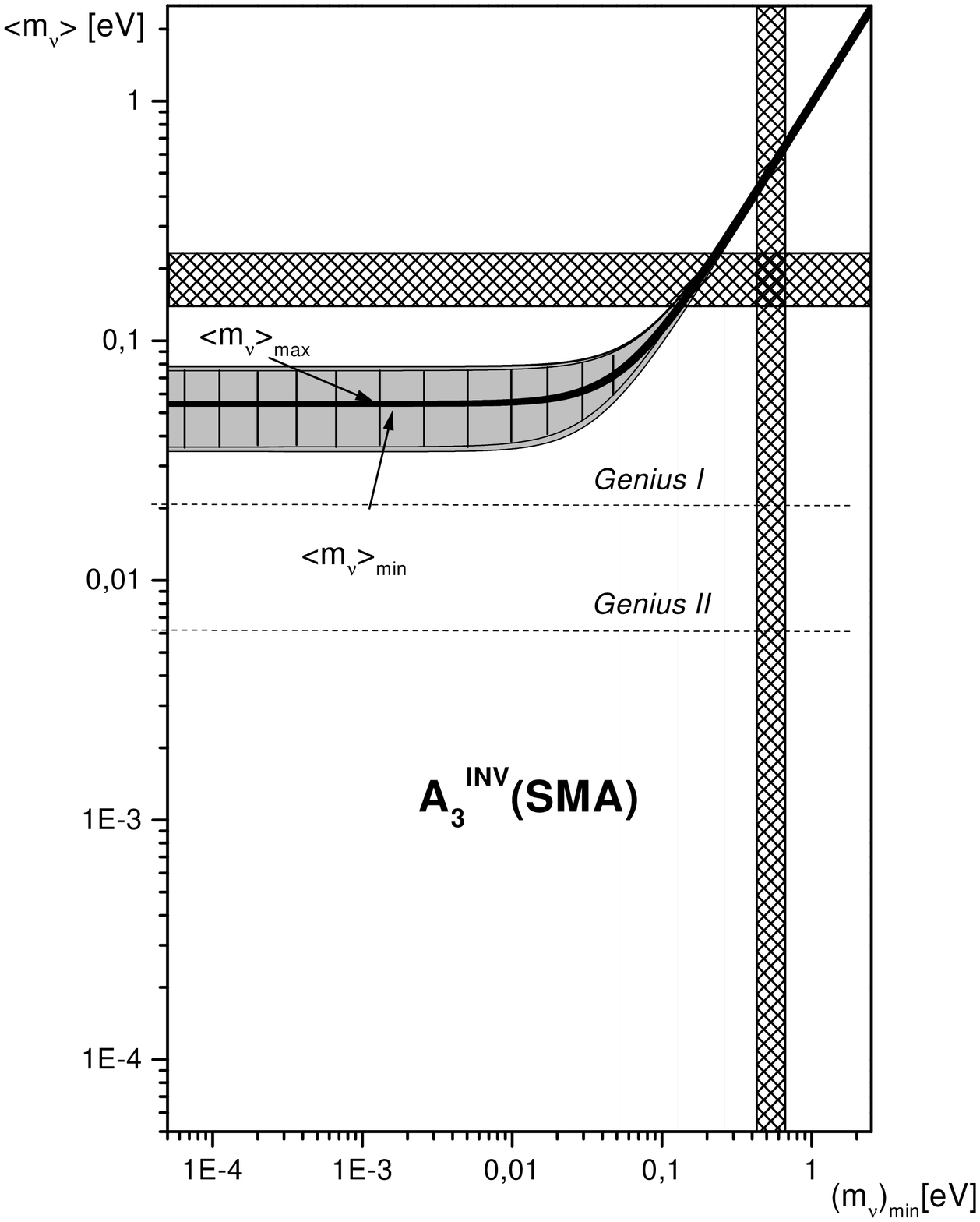, scale=0.6, height=13cm, width=15cm}
\caption{
The value of $\left< m_{ \nu} \right>_{max}$ and $\left< m_{ \nu}
\right>_{min}$  as function of $(m_{ \nu})_{min}$
for SMA MSW scenario and $A_{3}^{inv}$ neutrino mass scheme.
The shaded and hashed regions correspond to the allowed ranges of
neutrino oscillation parameters (Table~1)
for $\left< m_{ \nu} \right>_{min}$ and $\left< m_{ \nu} \right>_{max}$,
respectively.
The solid lines correspond to the best fit values of neutrino
oscillation parameters.}
\label{smainv}
\end{center}
\end{figure}

As $\theta_{13}$ is small (Eq.~\ref{mm}), for degenerate neutrino masses
 $\left<m_{\nu} \right>_{min}$  depends crucially on  $ \epsilon$.\\
 If $ \epsilon \rightarrow 1$ (which is realized for SMA MSW solution)
\begin{eqnarray}
\label{mnucos}
\left<m_{\nu} \right>_{min} \approx (m_{\nu})_{min}cos2 \theta_{13},
\end{eqnarray}
and the spread of the region $ \Delta \left<m_{\nu} \right>=
\left<m_{\nu} \right>_{max}- \left<m_{\nu} \right>_{min}$ for a
given value of $(m_{\nu})_{min}$ is small
(see Figs.~\ref{sma},\ref{smainv})
\begin{eqnarray}
\label{frac1}
\frac{\Delta \left<m_{\nu} \right>}{(m_{\nu})_{min}} =2 \ sin^{2}
\theta_{13}<0.08.
\end{eqnarray}

\begin{figure}[t]
\begin{center}

\epsfig{figure=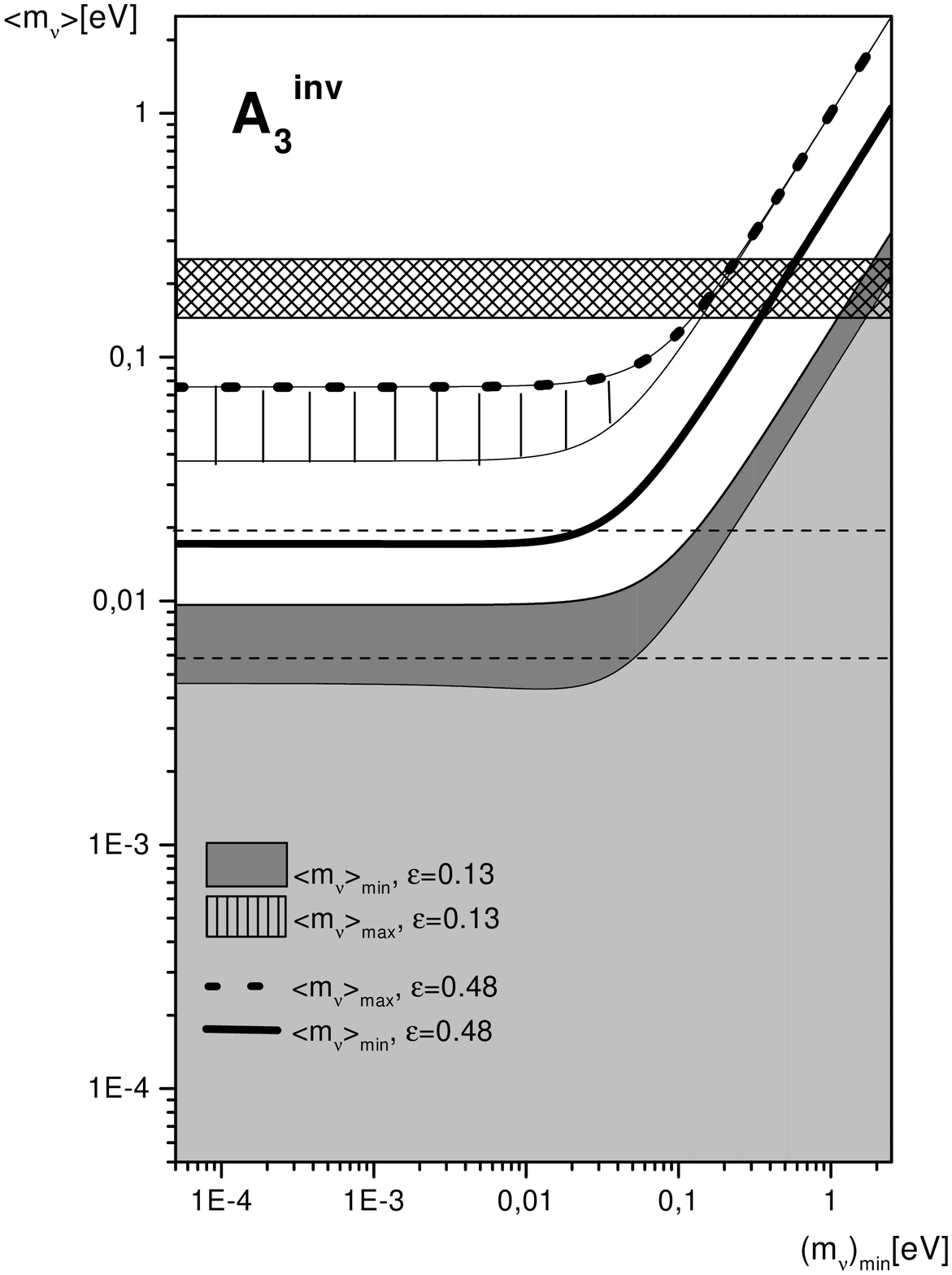, scale=0.6, height=13cm, width=15cm }

\caption{
The value of $\left< m_{ \nu} \right>_{max}$ and $\left< m_{ \nu}
\right>_{min}$  as function of $(m_{ \nu})_{min}$
for LMA (LOW-QVO)  MSW scenario and $A_{3}^{inv}$ neutrino mass scheme.
Shaded area shows $\left< m_{ \nu} \right>_{min}$ for $ \epsilon=0.13$
and $ \delta m_{atm}^2$,$ |U_{e3}|^{2}$,$\delta m^2_{sol}$ parameters in
a full range of their present possible values
(see Table~1). Hashed area shows the same for  $\left< m_{ \nu}
\right>_{max}$.
Horizontal band corresponds to  $\left< m_{ \nu} \right>$ as planed by
GENIUS I (with some anticipated error).
The thick solid (dashed) line correspond to   $\left< m_{ \nu}
\right>_{min} ($ $\left< m_{ \nu} \right>_{max}$)
and $\epsilon=0.48$. This time neutrino oscillation parameters are taken
with their best values (Table~1).}
\label{a3inv}
\end{center}
\end{figure}

For the LMA and LOW-QVO solutions of the solar neutrino problem
$\epsilon \ll 1$. In this case strong cancellations in $\left<m_{\nu}
\right>_{min}$ occur and values $\left<m_{\nu} \right>_{min} \approx 0$,
even for large $(m_{\nu})_{min}$, are not excluded.
Also the spread of the region $\Delta \left<m_{\nu} \right>$ is
substantial
\begin{eqnarray}
\label{epsilon2}
\frac{\Delta \left<m_{\nu} \right>}{(m_{\nu})_{min}} \approx
\frac{\left<m_{\nu} \right>_{max}}{(m_{\nu})_{min}} \rightarrow 1.
\end{eqnarray}
The relations mentioned above are depicted in
Figs.~\ref{a3}-\ref{a3inv}.

Now we will discuss the results gathered in Figs.~\ref{sma}-\ref{a3inv}
in more details.

\subsubsection {Majorana neutrinos and SMA MSW}

\begin{figure}[t]
\begin{center}
\epsfig{figure=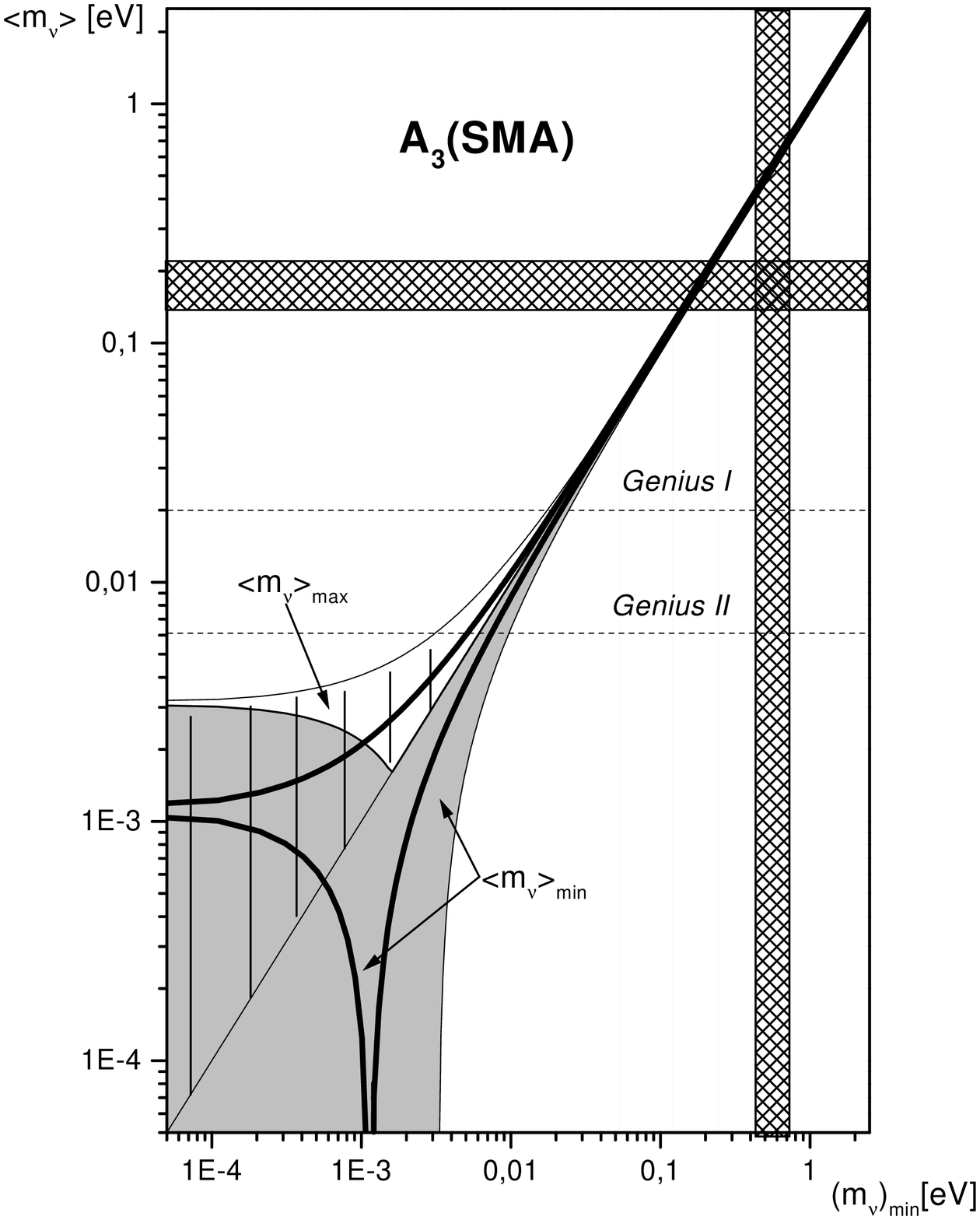, scale=0.6, height=13cm, width=15cm}
\caption{The value of $\left< m_{ \nu} \right>_{max}$ and $\left< m_{
\nu} \right>_{min}$  as function of $(m_{ \nu})_{min}$
for SMA MSW scenario and $A_{3}$ neutrino mass scheme.
The shaded and hashed regions correspond to the allowed ranges of
neutrino oscillation parameters (Table~1)
for $\left< m_{ \nu} \right>_{min}$ and $\left< m_{ \nu} \right>_{max}$,
respectively.
The solid lines correspond to the best fit values of neutrino
oscillation parameters.
The experimental bound on $\left< m_{ \nu} \right>$ planed by GENIUS I
and GENIUS II are depicted (dashed, horizontal lines).
The vertical band correspond to the possible range of $(m_{\nu})_{min}$
determined by the  tritium $\beta$ decay experiment.}
\label{sma}
\end{center}
\end{figure}

Figs.~\ref{sma},\ref{smainv} show the allowed $\left<m_{\nu} \right>$
range for the SMA MSW solution.
The solid lines represent  $\left<m_{\nu} \right>_{max}$ and
$\left<m_{\nu} \right>_{min}$ for the best fit parameters.
The shaded and hashed regions correspond to  uncertainties of
the oscillation parameters
(Table ~1)
for  $\left<m_{\nu} \right>_{min}$ and $\left<m_{\nu}\right>_{max}$,
respectively.

In the $A_3$ scheme (Fig.~\ref{sma})
the present bound on $\left<m_{\nu}\right>$ (Eq.~\ref{pres}) implies the
largest value of   $(m_{\nu})_{min}$
\begin{eqnarray}
\label{.2}
(m_{\nu})_{min}<0.2 \ eV,
\end{eqnarray}
and from Eqs.~\ref{m3},\ref{.04}
\begin{eqnarray}
\label{.21}
0.04 \leq (m_{\nu})_{max} \leq 0.21 \ eV.
\end{eqnarray}
Future bounds on $\left<m_{\nu} \right>_{exp}$, inferred from $(\beta
\beta)_{0 \nu}$ experiments,
have chance to give a stringent limit on neutrino masses. For example
\begin{enumerate}
\item if  $ \left<m_{\nu} \right><0.02 \ eV$ (GENIUS 1t), then
\begin{eqnarray}
\label{RRR}
(m_{\nu})_{min}<0.024 \ eV \Rightarrow  0.04 \ eV \leq  (m_{\nu})_{max}
\leq 0.063 \ eV,
\end{eqnarray}
\item if  $ \left<m_{\nu} \right><0.006 \ eV$ (GENIUS 10t), then
\begin{eqnarray}
\label{RRR1}
(m_{\nu})_{min} <0.01 \ eV \Rightarrow  0.04 \ eV \leq (m_{\nu})_{max}
\leq 0.059 \ eV.
\end{eqnarray}
\end{enumerate}
%In both cases we use present values of $\delta m^{2}_{atm}$ from Table
%~1, which probably will be much better
%when GENIUS results appear to be..............

It is also possible to find Majorana neutrino masses if the $(\beta
\beta)_{0 \nu}$ decay is observed and a value
$\left<m_{\nu} \right>_{exp} \in (0.006 \div 0.2) \ eV$ is inferred. We
can see from Fig.~\ref{sma}  that the range $ \Delta \left<m_{\nu}
\right>$
is up to $(m_{\nu})_{min} \approx 0.015 \ eV$ reasonably narrow and
the knowledge of $\left<m_{\nu} \right>_{exp}$
gives a  chance to determine  $(m_{\nu})_{min}$ with a good  precision.
For instance, if $ \left<m_{\nu} \right>_{exp} \approx 0.006 \ eV$ then
$(m_{\nu})_{min} \sim (3 \div 10) \times 10^{-3} \ eV$, the determination of
smaller values of $(m_{\nu})_{min} \ll  \sqrt{ \delta m_{atm}^{2}}$ for the
hierarchical mass spectrum is impossible with the present oscillation
parameters uncertainties.

In the case of $A_{3}^{inv}$ scheme  (Fig.~\ref{smainv}), the shaded and
hashed regions which describe the uncertainty in the determination of
$\left<m_{\nu} \right>_{min}$ and
$ \left<m_{\nu} \right>_{max}$
are almost identical and narrow. From the present limit  on
$\left<m_{\nu} \right>$ (Eq.~\ref{pres}) it follows that
\begin{eqnarray}
\label{R}
\left< m_{\nu} \right> \ < \ 0.2 \ eV \Rightarrow   (m_{\nu})_{min} \ <
\  0.22 \ eV.
\end{eqnarray}
Future bounds on $\left<m_{\nu} \right>$ up to $ \left<m_{\nu}
\right>_{exp} \approx 0.04 \ eV$,
still give the upper limit on $(m_{\nu})_{min}$. If the bound on
$\left<m_{\nu} \right>$ is smaller (GENIUS I) the scheme $A_{3}^{inv}$
is excluded for Majorana neutrinos.

The $A_3$ scheme can not be excluded in this way, even for very small $
\left<m_{\nu} \right>_{exp}$.
However, we can also consider a hypothetical situation. 
Let us imagine that the $^{3}_{1}He$ decay measurement
gives the mass of  $(m_{\nu})_{min}$ in the region (0.4$ \div$ 0.7) eV
(see Fig.~\ref{sma}). At the same time
the GENIUS I experiment will moves the limit to $\left<m_{\nu} \right><0.02 \
eV$. So, from the second information it follows, that
$(m_{\nu})_{min}$ is smaller than 0.024 eV (Eq.~\ref{RRR}), which is in
evident conflict with the $^{3}_{1}H$ decay measurements. There is only
one obvious conclusion in this case. Neutrinos cannot be Majorana
particles, they must have a Dirac nature.

We can see that the SMA MSW solution gives a crucial information about
the Majorana neutrino mass spectrum independently
of  future $(\beta \beta)_{0 \nu}$ experiments giving a bound, or finding a
finite value for $\left<m_{\nu} \right>_{exp}$.
Unfortunately, among the four possible solutions of the solar neutrino
problem, the SMA gives presently the worst goodness of fit \cite{3n}.

\subsubsection{Majorana neutrinos and LMA, LOW-QVO solutions: $A_3 $
scheme.}

%{\bf In this case the angle $\theta_{12}$ can take values including
%$\theta _{12}= \frac{ \pi }{4}$ (Table ~1) with $95 \%$ CL.}

\begin{figure}[t]
\begin{center}

\epsfig{figure=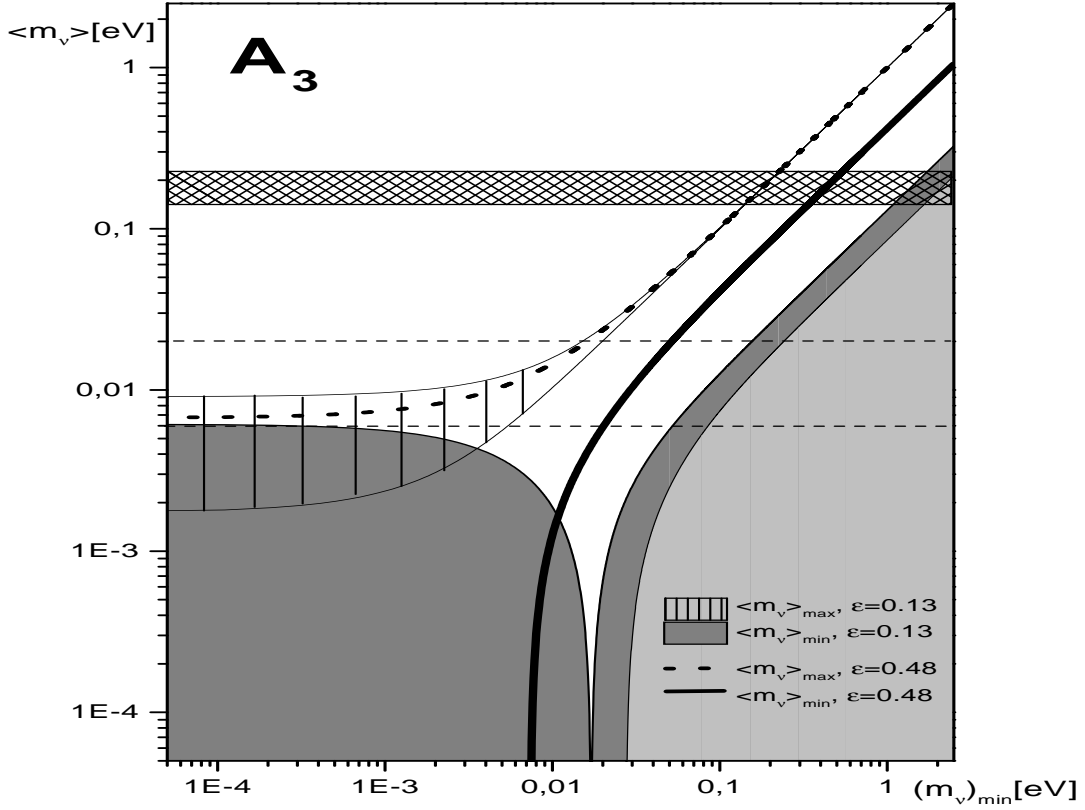, scale=0.6, height=13cm, width=15cm }

\caption{
The value of $\left< m_{ \nu} \right>_{max}$ and $\left< m_{ \nu}
\right>_{min}$  as function of $(m_{ \nu})_{min}$
for LMA (LOW-QVO)  MSW scenario and $A_{3}$ neutrino mass scheme.
Shaded areas shows $\left< m_{ \nu} \right>_{min}$ for $ \epsilon=0.13$
and $ \delta m_{atm}^2$,$ |U_{e3}|^{2}$,$\delta m^2_{sol}$ parameters in
a full range of their present possible values
without error of $\epsilon$ (dark shaded) and with this error (light
shaded)
(see Table~1). Hashed region shows   $\left< m_{ \nu} \right>_{max}$
with
$\delta m_{atm}^2$,$ |U_{e3}|^{2}$,$\delta m^2_{sol}$ parameters also in
a full range of their present possible values.
Horizontal band corresponds to  $\left< m_{ \nu} \right>$ as planed by
GENIUS I (with some anticipated error).
The thick solid (dashed) line correspond to   $\left< m_{ \nu}
\right>_{min} ($ $\left< m_{ \nu} \right>_{max}$)
and $\epsilon=0.48$. This time neutrino oscillation parameters are taken
with their best values (Table~1).}
%and vertical one to the possible value of $m_{ \beta}$ in future
%$^{3}_{1}H$ decay experiment.}
\label{a3}
\end{center}
\end{figure}

\begin{figure}[t]
\begin{center}

\epsfig{figure=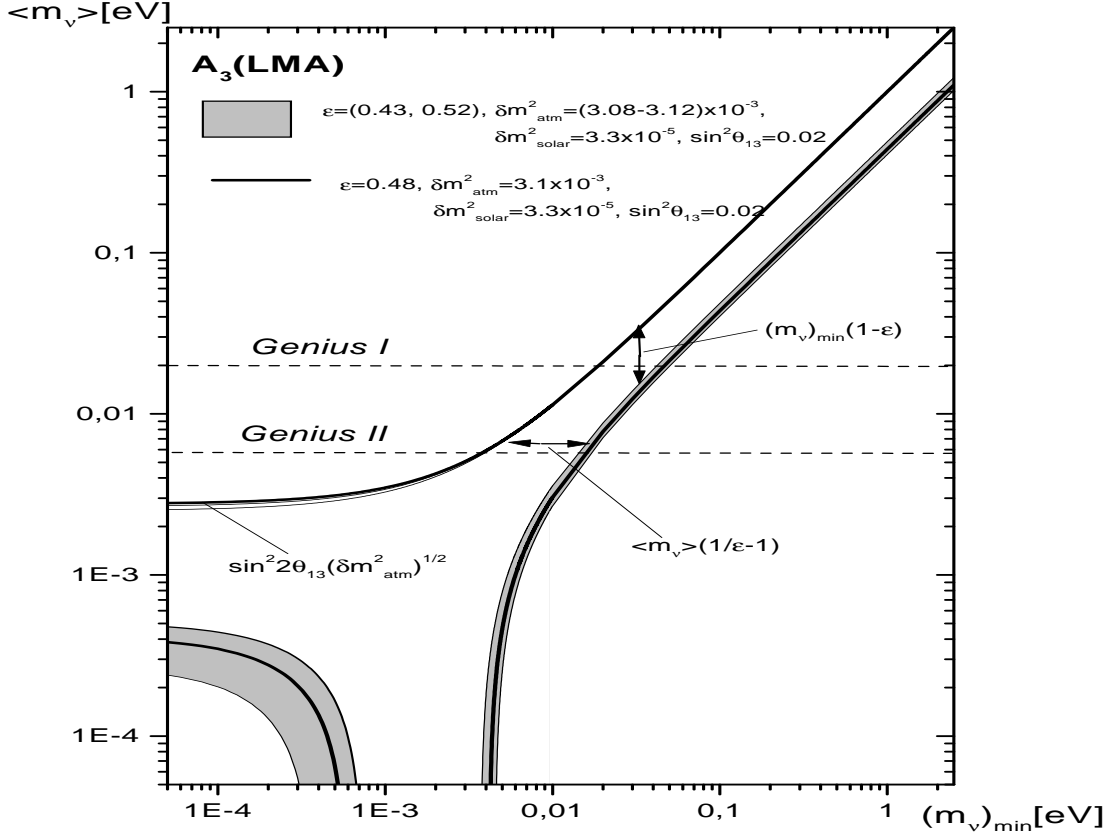, scale=0.6, height=13cm, width=15cm}

\caption{The value of $\left< m_{ \nu} \right>_{max}$ and $\left< m_{
\nu} \right>_{min}$  as function of $(m_{ \nu})_{min}$
for LMA MSW scenario and $A_{3}$ neutrino mass scheme, $ \epsilon=0.48$.
This time, opposite to the case
of Figs.~\ref{a3},\ref{a3inv} the anticipated error of 10 \% in future
$\sin^2{2 \Theta_{sol}}$ determination
 is included, $\epsilon \simeq \left( 0.43 \div 0.52 \right )$. Expected
improvement
in $ \delta m_{atm}^2$,$ |U_{e3}|^{2}$,$\delta m^2_{sol}$ parameters
determination is also taken into account.
See the last column in Table~1 and the text for details.
}
\label{futa3}
\end{center}
\end{figure}

In Figs.~\ref{a3},\ref{a3inv}  regions of $\Delta \left<m_{\nu} \right>$ as
function of $(m_{\nu})_{min}$ for
various  $ \epsilon$ values are shown. $\epsilon=0.13$ corresponds to value
of $\sin^2{2 \Theta_{sol}}=0.98$ ($\tan^2{\Theta_{sol}}=0.77$) 
while  $\epsilon=0.48$ ($\sin^2{2 \Theta_{sol}}=0.77$) is 
present best fit
value (see Table~1).
The dark shaded area shows the influence of the  uncertainties connected
to the present neutrino oscillation parameters' determination ($\delta
m^2_{21}, \delta m^2_{32},\sin^2{\Theta_{13}}$ in Table~1)
on  $\left<m_{\nu} \right>_{min}$ for  $\epsilon=0.13$. At 95 \% c.l.
$\epsilon=0$ is possible
and $\left<m_{\nu} \right>_{min}$
reaches zero also for higher values of $(m_{\nu})_{min}$  (light shaded
region).
The hashed region describes the influence of the present oscillation
parameters' uncertainties
 on  $\left<m_{\nu} \right>_{max}$ (once more with constant $\epsilon$).

Let us now consider two situations, first when a future new bound on
$\left<m_{ \nu } \right>_{exp} \ < \
\kappa \in \ (0.2 \div 0.006)$ eV is obtained and
second when some value  $\left<m_{ \nu } \right>_{exp} = \kappa +
\Delta \kappa$ is definitely found.

\underline{No signal for  $(\beta \beta)_{0 \nu}$  in future}
%Neutrinoless double beta decay not found $\left<m_{ \nu } \right>_{exp}
%\ < \ \kappa$} \\

The information on the Majorana neutrino masses can be inferred only if
$\left<m_{ \nu } \right>_{min} \neq 0$ which is
equivalent to
$\epsilon \ > \tan^2 \theta_{13}$. In this case the condition  $\langle
m_{\nu} \rangle_{min} < \kappa $
gives nontrivial bounds on  $(m_{\nu})_{min}$  (Eq.~\ref{sc})

\begin{eqnarray}
\label{kappa2}
(m_{\nu})_{min} < \frac{ \kappa }{ \cos^{2} \theta
_{13}(\epsilon-\tan^{2} \theta _{13})}.
\end{eqnarray}
For small values of $ \epsilon$ such a bound can be less restrictive
than the one obtained from the $^3_{1} H$ decay.
We would like to concentrate on the possibility that the future solar
neutrino experiments give the value
$ \epsilon \gg  \tan^{2} \theta _{13}$.  Then
\begin{equation}
 (m_{\nu})_{min}  \leq  \frac{ \kappa}{ \epsilon} .
\end{equation}

If, $ \epsilon =0.13(0.48)$ (Fig~\ref{a3}) then present experimental
bound on $\left<m_{\nu} \right>$
gives the upper limit on $(m_{ \nu})_{min}$,
\begin{eqnarray}
\label{mnimin}
(m_{\nu})_{min} < 2.2(0.4) \ eV ,
\end{eqnarray}
which is better than the tritium $\beta$ decay (Eq.~\ref{2.2}) for a
larger value of $\epsilon$.
If $\left<m_{\nu} \right>_{exp}<0.006 \ eV$,
 then $(m_{\nu})_{min}<0.022 \ eV$ ($\epsilon=0.48$) and
$(m_{\nu})_{min}<0.085 \ eV$ ( $\epsilon=0.13$).

So, if $( \beta \beta)_{0 \nu}$ is not found, we do not know if
neutrinos are Majorana particles. If they indeed are
then limits on neutrino masses can be found which are improving
for increasing $\epsilon$.

\newpage 

\underline{Positive signal for $\langle \beta \beta \rangle_{0 \nu}$ in
future}\\

From the   $\langle \beta \beta \rangle_{0 \nu}$  measurement we know
that
\begin{eqnarray}
\label{kappa3}
\left<m_{\nu} \right>_{exp} = \kappa \pm \Delta \kappa .
\end{eqnarray}

Obviously, in this case consistency requires

\begin{eqnarray}
\label{kappa4}
\left<m_{\nu} \right>_{max} > \kappa - \Delta \kappa ,
\end{eqnarray}

from which interesting informations on the Majorana neutrino masses can be
obtained, even for small values of $\epsilon \to 0$.
This situation  was already considered in  ~\cite{many,last}, so we will
not analyze it in detail.
We only add some numbers which follow from Fig.~\ref{a3}.
The measured values of $\left<m_{\nu} \right>_{exp} \geq 0.01 \ eV$ are able
to bound $(m_{\nu})_{min}$ from below, e.g.,
$\left<m_{\nu} \right>_{exp} \approx 0.03 \ eV$ gives $(m_{\nu})_{min}
\geq 0.025 \ eV$.
Depending on the measured values of $\left<m_{ \nu} \right>_{exp}$,
various mass schemes can be excluded or allowed. For instance,
if $\left<m_{\nu} \right>_{exp} \leq 0.01 \ eV$ then  $(m_{\nu})_{min}$
can be very small
and the hierarchical mass spectrum is allowed. For  $\left<m_{\nu} \right>
\geq  0.03 \ eV$ only the degenerate spectrum will be acceptable.

A new situation occurs if $\left<m_{\nu} \right>_{exp} \geq 0.01 \
eV$ and $\epsilon \gg \tan^2 \theta_{12}$.
Then from
$\left<m_{\nu} \right>_{min}$ an upper limit on $(m_{\nu})_{min}$
can also be found. It means that a
finite range of possible values of $(m_{\nu})_{min}$ can be derived
\begin{eqnarray}
\label{range1}
(m_{\nu})_{min} \in ( (m_{\nu})_{min}^{min},(m_{\nu})_{min}^{max} ),
\end{eqnarray}
where
\begin{eqnarray}
\label{minmin}
(m_{\nu})_{min}^{min}&=& \kappa - \Delta \kappa \nonumber \\
(m_{\nu})_{min}^{max}&=& \left(  \kappa + \Delta \kappa  \right)
\frac{1}{(1+ \epsilon ) \cos^{2} \theta _{13}-1}.
\end{eqnarray}
The uncertainty of the $(m_{\nu})_{min}$ determination
\begin{eqnarray}
\label{delta}
\Delta (m_{\nu})_{min}=(m_{\nu})_{min}^{max}-(m_{\nu})_{min}^{min}
\end{eqnarray}
decreases  with increasing $\epsilon$

\begin{eqnarray}
\label{ul}
\frac{\Delta (m_{\nu})_{min}}{\kappa}&=& \frac{2-(1+ \epsilon ) \cos^{2}
\theta _{13}}{(1+ \epsilon) \cos^2 \theta _{13}-1} +
\frac{\Delta \kappa}{\kappa} \frac{(1+ \epsilon ) \cos^{2} \theta
_{13}}{(1+ \epsilon) \cos^2 \theta _{13}-1}  \nonumber \\
&  \stackrel{\theta_{13} \rightarrow 0}{\rightarrow} &  \ \left(
\frac{1}{ \epsilon }-1 \right) + \frac{\Delta \kappa}{\kappa}
\left(  \frac{1}{ \epsilon }+1 \right).
\end{eqnarray}

From Fig. ~\ref{a3} and for $\epsilon =0.13$ we can read (neglecting
$\Delta \kappa$)
\begin{eqnarray}
\label{for2}
\left<m_{\nu} \right>_{exp}=0.2 \ eV  & \Rightarrow & (m_{ \nu})_{min}
\in (0.2 \div 2.2) \ eV, \\
\label{for1}
\left<m_{\nu} \right>_{exp}= 0.02 \ eV  & \Rightarrow & (m_{ \nu})_{min}
\in (0.012 \div 0.2) \ eV.
\end{eqnarray}

% where experimental uncertainties of oscillation parameters are taken
%into account.

Moreover, if $\left<m_{\nu} \right>_{exp} <0.01$ eV we can say only that
$(m_{\nu})_{min}<0.02 \div 0.05$ eV.

So,  for LMA and LOW-QVO solutions, the knowledge of  
$\left< m_{\nu}\right>_{exp}$
is able to restrict the Majorana neutrino mass spectrum, as long as
$\epsilon \geq \tan^2{\Theta_{13}}$.
The range of $ (m_{ \nu})_{min}$ (Eq.~\ref{delta}) depends on $\epsilon$
and $\Delta \kappa$ and is smaller for
larger values of $\epsilon$.

In Fig.~\ref{futa3} the region $(\left< m_{\nu} \right>_{min}, \left<
m_{\nu} \right> _{max})$ is shown for the case
$\epsilon =0.48$ ($\sin^2{2 \Theta_{sol}}=0.77$) where anticipated,
much smaller errors on $\delta m_{atm}^2$, $|U_{e3}|^2$ and $\delta
m_{sol}^2$
compared to the present ones are taken into account (see Table~1 for
details). The expected 10 \% error of $\sin^2{2 \Theta_{sol}}$ is included.
The uncertainty of
$ \left< m_{\nu} \right> _{max}$ is now  almost invisible. For $ \left<
m_{\nu} \right> _{min}$ two separated
regions of nonzero  $\left<m_{\nu} \right>$ are present. The light
shaded region as in Fig.~\ref{a3} does not
appear at all.
The shape of the $ \left< m_{\nu} \right>_{min}$ depends on the value of
$\epsilon$ and this presented in
Fig.~\ref{futa3} is typical of non negligible values of $\epsilon$. We can see
that for
the experimental  bound $\left<m_{\nu} \right>_{exp} < \kappa$ the upper
limit of  $ (m_{ \nu})_{min}$ can be easily
found. The same is  true of the case   $\left<m_{\nu} \right>_{exp} =
\kappa \pm \Delta \kappa$ where the range of
possible   $ (m_{ \nu})_{min}$  can be found again. There is only one
important modification. If the future bound
or the experimental value of  $\left<m_{\nu} \right>_{exp}$ will be
smaller than 0.001 eV then the mass $ (m_{ \nu})_{min}$
of Majorana neutrinos will be limited from below and, this time, also
from above. For instance, if
 $\left<m_{\nu} \right>_{exp}<2 \cdot 10^{-4}$ eV, then  $ (m_{
\nu})_{min} \in \left( 1.0 \cdot 10^{-4}\ \mbox{\rm eV},
5.0 \cdot 10^{-3}\ \mbox{\rm eV} \right)$.

For larger values of $\epsilon$ ($\epsilon \sim 0.5 )$ it can also
happened that the accepted range of
$(m_{\nu})_{min}$ found from tritium $\beta$ decay is in conflict
with a bound given by
$(\beta \beta)_{0 \nu}$ decay.  The situation is practically the same as in
the SMA case and the
conclusions concerning the neutrino nature are the same (see previous
disscusion).
From such a unique scenario follows that neutrinos are Dirac particles.

%For very small $\theta_{13}$ angle, better bound on mass of the
%lightest neutrino $(m_{\nu})_{min}$ is obtained
%if $\epsilon$ is larger.

Now we would like to comment on the CP phases.
The effect of two unknown  CP Majorana phases disappears if $\epsilon
\rightarrow 1$ (SMA).
So for large $\epsilon$ the information about the CP phases is lost. If,
however, $\epsilon$ is small
 ($\epsilon \sim 0.1 \div 0.5 )$ and  $\left< m_{\nu} \right>_{exp}=
\kappa \pm \Delta \kappa$, $m_{\beta}=\kappa' \pm \Delta \kappa'$
are found with a good precision, some insight into the CP symmetry is
possible.
Comparing both bands $ \left< m_{\nu} \right> \in ( \kappa - \Delta
\kappa , \kappa + \Delta \kappa)$
and $m_{\beta} \in ( \kappa' - \Delta \kappa' , \kappa' + \Delta
\kappa')$
with the $\left( \left< m_{\nu}\right>_{min},\left<m_{\nu}\right>_{max}
\right)$ region allowed by the oscillation data
is a check of internal consistency of the theory. With precise data the
crossing of the three regions can be used to
specify the values of the CP breaking Majorana phases
(Eqs.~\ref{basic},\ref{basic1}).
If the two bands $\left< m_{\nu} \right>$ and $m_{\beta}$ cross the
oscillation region near $ \left< m_{\nu}\right>_{max}$ then
two phases are equal $\phi_1=\phi_2 \approx n \pi$. This means that all
three Majorana neutrinos have the same CP parity
$ \eta _{CP}=+i$ and the symmetry is conserved. If the two bands cross
the oscillation region near $\left< m_{\nu} \right>_{min}$,
once more the CP symmetry is satisfied and
$\eta_{CP}(\nu_1)=-\eta_{CP}(\nu_2)=-\eta_{CP}(\nu_3)=i$.
Finally, if all three regions cross somewhere in between, the phases
$\phi_1$ and $\phi_{2}$ are nontrivial and the CP symmetry is broken. We
can also imagine the situation that all three regions do not cross in
the same place.
This would be a signal that the theory with three light Majorana
neutrinos is not consistent.\\
\subsubsection{Majorana masses and LMA, LOW-QVO solutions: $
A_{3}^{inv}$ scheme}

 The same analysis as before can be done for the $A_{3}^{inv}$ scheme.
For degenerate masses  ($(m_{ \nu})_{min} \geq 0.2 \ eV$), two functions
$\left<m_{\nu} \right>_{min}$
and $\left<m_{\nu} \right>_{max}$ are exactly the same as in the
$A_{3}$ scheme (Eqs.~\ref{4lin},~\ref{sc}).
So  conclusions concerning the determination of the Majorana neutrino masses
are the same.
The behavior of the functions $\left<m_{ \nu} \right>_{min(max)}$ is
different for small values of $(m_{ \nu})_{min}$.
As can be seen from Figs.~\ref{smainv},\ref{a3inv} $\left<m_{ \nu}
\right>_{min}$ never vanishes if $ \epsilon \neq 0$.
The minimal value of  $\left<m_{ \nu} \right>_{min}$is proportional to $
\epsilon $, namely
\begin{eqnarray}
\label{fmin}
\left<m_{ \nu} \right>_{min} \left[ (m_{\nu})_{min}\approx 0 \right]
\approx \epsilon \cos^{2} \theta _{13} \sqrt{ \delta m^{2}_{atm}}
\approx \epsilon \cdot 0.04.
\end{eqnarray}
The minimal value of $\left<m_{\nu} \right>_{max}$ does not depend on
$\epsilon$ and
\begin{eqnarray}
\label{gmin}
\left<m_{\nu} \right>_{max} \left[ (m_{\nu})_{min} \approx 0 \right]
\approx \cos^{2} \theta _{13}\sqrt{ \delta m^{2}_{atm}} \approx  0.08.
\end{eqnarray}
So, if in future the $(\beta \beta)_{0 \nu}$ decay gives a bound  $\left<m_{
\nu} \right>_{exp} < \kappa $, then the
scheme $A_{3}^{inv}$ has to be rejected (for Majorana neutrinos) or
neutrinos are Dirac particles when
\begin{eqnarray}
\label{epssin}
\epsilon \cos^{2} \theta_{13} \sqrt{ \delta m^{2}_{atm}} > \kappa.
\end{eqnarray}
If, on the other hand, a finite value of ${\langle m_{\nu}
\rangle}_{exp}= \kappa \pm \Delta \kappa$ is found  then
three scenarios are possible
\begin{enumerate}
\item $\kappa - \Delta \kappa >
\cos^2 \theta_{13}\sqrt{\delta m^{2} _{atm}}$. The lightest neutrino
masses $(m_{\nu})_{min}$ can be bounded from below;
\item  $\epsilon \cos^{2} \theta ^{2} _{13} \sqrt{ \delta m^{2}_{atm}}
-  \Delta \kappa
< \kappa < \cos^{2} \theta_{13}\sqrt{\delta m_{atm}^{2}} +  \Delta
\kappa $.
The  mass $(m_{\nu})_{min}$ is weakly bounded to the region
$(m_{\nu})_{min} \leq 0.05$ eV; \newline
and finally
\item  $\kappa +  \Delta \kappa  < \epsilon \cos^2 \theta_{13}
\sqrt{\delta m_{atm}^2}$.
The scheme $A_3^{inv}$ is excluded.
\end{enumerate}

\section{Conclusions}

Atmospheric and solar neutrino experiments give strong evidence that
neutrinos have non zero masses and that they mix.
However, these experiments alone are not able to determine the absolute
neutrino masses.
All other terrestrial experiments are consistent with the assumption
that neutrinos are massless
\cite{apbgen,dm}.
Only tritium $\beta$  and  $( \beta \beta)_{0 \nu}$ decays are sensitive
to neutrino masses at the ${\cal{O}}(eV)$
level and the confirmation  of their existence
at this scale seems to be just around the corner. However, even there
only
the combination of neutrino masses can be determined.
 We have considered whether and how precisely the present and future
experimental data can determine the single absolute neutrino masses.
 With the present experimental precision we have found
\begin{eqnarray}
\label{mimj}
|m_{i}-m_{j}|&<&0.08 \ eV, \ \ \ \ , \ \ \ \ i,j=1,2,3 \nonumber \\
m_{i}&<&2.2 \ eV, \nonumber \\
max(m_{1},m_{2},m_{3})&>&0.04 \ eV.
\end{eqnarray}

In the future tritium beta decay altogether with oscillation experiments
are the best options to reconstruct the absolute values of
neutrino masses, independently of whether they are
Dirac or Majorana particles.
The relative error which comes from the uncertainty of the oscillation
parameters is very small and  has no influence on the neutrino
mass determination. The results depend uniquely on the precision to
which $m_{\beta}$ can be determined.
That is why this procedure is effective for neutrino masses above $\sim
0.2 \div 0.3$ eV.
 This will be a challenge for future experiments.

If neutrinos are Majorana particles additional information can be
inferred from the neutrinoless double beta decay, independently
if a non-zero $(\beta \beta)_{0 \nu}$ decay rate is found or not. There
is only one difference, in the second case we have
no experimental confirmation that they really have the Majorana nature. The
precision depends on the solution of the solar neutrino problem.
Solutions with smaller  $sin^2 2 \theta_{solar}$ are  better for
the neutrino mass determination.
For the SMA solution ($\epsilon \simeq 1$) we found
\begin{eqnarray}
\label{rr}
0 \leq &min(m_{i})& \leq 0.2 \ eV, \nonumber \\
0.04 \ eV\leq &max(m_{i})& \leq 0.21 \ eV,
\end{eqnarray}
which is a much stringer bound than Eq.~\ref{mimj}.
 If the SMA solution is confirmed by the future data, the next generation of the
$(\beta \beta)_{0 \nu}$ experiments has a good chance to find neutrino
masses as small as $(m_{\nu})_{min} \approx 0.015 \ eV$. Unfortunately,
the SMA scenario is presently
not a favored solution of the solar neutrino problem.

The neutrino mass determination in the case of  the solar neutrino
anomaly with small $\epsilon$  (LMA, LOW-QVO)
is more complicated. First of all  $\left<m_{ \nu } \right>_{min} =0$
for $\epsilon \leq \tan^2{\Theta_{13}}$
and the upper  limit on $ (m_{\nu})_{min}$
can not be obtained (the lower limit is given).
If, however, $\epsilon \ > \tan^2 \theta_{13}$ then the derivation of some
useful upper bounds is possible.
We have found the analytical bound on $(m_{\nu})_{min}$
(Eq.~\ref{kappa2}) given by the experimental
limit on $\left< m_{\nu} \right>_{exp}$ and the parameter $\epsilon$.
We have also found the uncertainty in the  $(m_{\nu})_{min}$ determination
$\Delta (m_{\nu})_{min}$ as function of
$\left<m_{\nu}\right>_{exp}$ and the two oscillation parameters
$\epsilon$,$\theta_{13}$.
%In numerical analysis present errors in neutrino parameters
%determination have been taken into account.

It can happen in future that the discovery of $(m_{\nu})_{min}$ from the
tritium $\beta$ decay
will be in conflict with the bound on $(m_{\nu})_{min}$ derived from the
$(\beta \beta)_{0 \nu}$ decay.
This is the unique situation where the Dirac character of neutrinos could be
confirmed.

For smaller values of $\epsilon$ and a good experimental precision to
which   $\left<m_{ \nu } \right>$
and $m_\beta$ can be determined, some insight into CP symmetry violation
or CP eigenvalues of neutrinos is possible.

{\bf Acknowledgments}
This work was supported by the Polish Committee for Scientific Research
under Grant No. 2P03B05418 and 5P03B08921.
M. C. would like to thank the Alexander von Humboldt Foundation for
fellowship.

\newpage

\begin{table}
{\scriptsize
\begin{tabular}{|c|c|c|c|c|c|}
\hline
\multicolumn{2}{|c|}{}
 & min. & best fit & max. & future improvements \\
\hline
\multicolumn{2}{|c|}{} &&&& \\
\multicolumn{2}{|c|}{$\tan^2{\Theta_{13}}$}
& 0 & 0.005 & 0.055 & $| \Delta \Theta_{13} | \sim  10^{-2}$
\cite{opera} \\
\multicolumn{2}{|c|}{} &  &  &  & $\;\;\;\;\;\;\;\;\;\;\; \sim  10^{-4}
\cite{era}$ \\
\multicolumn{2}{|c|}{} & &  & & \\
\hline
\multicolumn{2}{|c|}{} &&&& \\
\multicolumn{2}{|c|}{ $ \delta m^2_{32} \;[\times 10^3\;eV^2]$} & 1.4
& 3.1 & 6.1 &  $| \Delta ( \delta m^2_{23}) | \sim 10\%$ acc.
\cite{opera}\\
\multicolumn{2}{|c|}{} &  &  &  & $\;\;\;\;\;\;\;\;\;\;\; \sim  1 \%$
\cite{fact}\\
\multicolumn{2}{|c|}{} & &  & & \\
\hline
\multicolumn{2}{|c|}{} &&&& \\
\multicolumn{2}{|c|}{$\tan^2{\Theta_{23}}$} & 0.39 & 1.4 & 3.0 &
 $| \Delta ( \sin^2{ 2 \Theta_{23}}) | \sim 5 \%$ acc. \cite{opera}  \\
\multicolumn{2}{|c|}{} &  &  &  & $\;\;\;\;\;\;\;\;\;\;\; \sim  1 \% $
\cite{era} \\
\multicolumn{2}{|c|}{} & &  & & \\
\hline
&&&&& \\
 & LMA $\times10^{5}$ & $\sim 1.6$
& 3.3& $\sim 20$ & $| \Delta ( \delta m^2_{21}) | \sim 10\%$ acc.
\cite{opera}
\\
&&&&& \\
$ \delta m^2_{21}\;[eV^2]$ & LOW  $\times10^{8}$ &  $\sim 0.08$ & 9.6 &
$\sim 30 $ & \\
& SMA  $\times10^{6}$ &  $\sim 4$ & 5.1 &  $\sim 9 $ & \\
&&&&& \\
\hline
&&&&& \\
 & LMA  & 0.2
& 0.36 & $\sim 1$ & $| \Delta ( \sin^2{ 2 \Theta_{12}}) | \sim 0.1$
 \cite{opera} \\
& & & &  & \\
&&&&& \\
$ \tan^2{\Theta_{12}}$ & LOW-QVO& $ 0.2$ & 0.58 &  $3 $ & \\
& & & & & \\
& SMA   &  $\sim 10^{-4}$ & $6.8 \times10^{-4}$ &  $\sim 2 \times10^{-3}
$ & \\
& & & & & \\
\hline
\end{tabular}
\caption{
The allowed ranges of neutrino parameters from
global analysis
%\cite{3n}
altogether with expected  future improvements (taken from M.C.
Gonzales-Garcia et al. in \cite{3n}).
In three central  columns minimum and  maximum are given at 90 \% c.l.
%\cite{3n}
Future improvements on parameters are mainly connected to
LMA MSW solutions
%\cite{kam}
and accelerator physics
(MINOS, ICARUS, OPERA  projects, $\nu$ factories).
}
}
\label{first}
\end{table}

\newpage

\newpage

\newpage

\newpage

\newpage

\begin{thebibliography}{99}

\bibitem{fermi} E. Fermi, Z. Phys. {\bf 88} (1934) 161.

\bibitem{furry} W. H. Furry, Phys. Rev. {\bf 56} (1939) 1184.
%\bibitem{review} see e.g.: S. M. Bilenky, C. Giunti,
W. Grimus, Prog. Part. Nucl. Phys. {\bf 43} (1999) 1.

\bibitem{davis}

K.~Lande, P.~Wildenhain, R.~Corey, M.~Foygel and J.~Distel,
%``Measurements of the B-8 and Be-7 nu/e fluxes from the sun with a
%Cerenkov triggered radiochemical neutrino detector,''
Nucl.\ Phys.\ Proc.\ Suppl.\  {\bf 91} (2001) 50;
E.~Bellotti,
%``First results from GNO,''
Nucl.\ Phys.\ Proc.\ Suppl.\  {\bf 91} (2001) 44;
V.~N.~Gavrin  [SAGE Collaboration],
%``Solar neutrino results from SAGE,''
Nucl.\ Phys.\ Proc.\ Suppl.\  {\bf 91} (2001) 36;
Y.~Fukuda {\it et al.}  [Kamiokande Collaboration],
%``Solar neutrino data covering solar cycle 22,''
Phys.\ Rev.\ Lett.\  {\bf 77} (1996) 1683;
W.\ Hampel {\em et al.},
                Phys.\ Lett.\ B {\bf 447}, 127 (1999);
Q.~R.~Ahmad {\it et al.}  [SNO Collaboration],
%``Measurement of the charged of current interactions produced by B-8
%solar neutrinos at the Sudbury Neutrino Observatory,''
nucl-ex/0106015.
\bibitem{fukuda}
S.~Fukuda {\it et al.}  [Super-Kamiokande Collaboration],
%``Constraints on neutrino oscillations using 1258 days of
%Super-Kamiokande solar neutrino data,''
Phys.\ Rev.\ Lett.\  {\bf 86} (2001) 5656.

\bibitem{athanassopoulos}
A.~Aguilar {\it et al.}  [LSND Collaboration],
%``Evidence for neutrino oscillations from the observation of  anti-nu/e
%appearance in a anti-nu/mu beam,''
hep-ex/0104049.

\bibitem{pakvasa}  S.~Pakvasa,
%``Neutrino anomalies without oscillations,''
Pramana {\bf 54} (2000) 65.

\bibitem{lipari} P.~Lipari and M.~Lusignoli,
%``Phenomenology of atmospheric neutrinos,''
hep-ph/9905229;
M.~Lusignoli,
%``Non-standard neutrino properties,''
Nucl.\ Phys.\ Proc.\ Suppl.\  {\bf 100} (2001) 250.

\bibitem{cmb}
J.R. Primack, astro-ph/0007187.
%\bibitem{eecr} cosmic rays
\bibitem{zburst}
 T.~J.~Weiler,
%``Cosmic ray neutrino annihilation on relic neutrinos revisited:  A
%mechanism for generating air showers above tcut-off,''
Astropart.\ Phys.\  {\bf 11} (1999) 303;
D.~Fargion, B.~Mele and A.~Salis,
%``Ultrahigh energy neutrino scattering onto relic light neutrinos in
%galactic halo as a possible source of highest energy erays,''
Astrophys.\ J.\  {\bf 517} (1999) 725;
T.~Weiler,
%``Resonant Absorption Of Cosmic Ray Neutrinos By The Relic Neutrino
%Background,''
Phys.\ Rev.\ Lett.\  {\bf 49} (1982) 234.

\bibitem{weiler}  H.~Pas and T.~J.~Weiler,
%``Absolute neutrino mass determination,''
Phys.\ Rev.\ D {\bf 63} (2001) 113015.

\bibitem{apbgen}  M. Czakon, J. Gluza, J. Studnik, M. Zra{\l}ek,
Acta Phys. Pol. {\bf B31} (2000) 1365.

\bibitem{notcl} S.J. Yellin, hep-ex/9902012;
A.~D.~Santo,
%``An experimentalist's view of neutrino oscillations,''
hep-ex/0106089.



\bibitem{sn1987}C.~Lunardini and A.~Y.~Smirnov,
%``Neutrinos from SN1987A, earth matter effects and the LMA solution of
%the solar neutrino problem,''
Phys.\ Rev.\ D {\bf 63} (2001) 073009;
H.~Minakata and H.~Nunokawa,
%``Inverted hierarchy of neutrino masses disfavored by supernova
%1987A,''
Phys.\ Lett.\ B {\bf 504} (2001) 301.

\bibitem{many}S.~M.~Bilenky, S.~Pascoli and S.~T.~Petcov,
%``Majorana neutrinos, neutrino mass spectrum, CP-violation and
%neutrinoless double beta-decay. I: The three-neutrino mixing case,''
hep-ph/0102265; S.~M.~Bilenky, S.~Pascoli and S.~T.~Petcov,
%``Majorana neutrinos, neutrino mass spectrum, CP-violation and  neutrinoless double beta-decay. II: Mixing of four neutrinos,''
arXiv:hep-ph/0104218.
%%CITATION = HEP-PH 0104218;%%



\bibitem{last} H.~V.~Klapdor-Kleingrothaus, H.~Pas and A.~Y.~Smirnov,
%``Neutrino mass spectrum and neutrinoless double beta decay,''
Phys.\ Rev.\ D {\bf 63} (2001) 073005.
%M.~Czakon, M.~Zralek and J.~Gluza,
%Acta Phys.\ Polon.\  {\bf B30} (1999) 3121



\bibitem{16a}
 S.T. Petcov, A. Y. Smirnov, Phys. Lett. {\bf B322} (1994) 109; S.M.
Bilenky, A. Bottino,
C. Giunti, C. Kim, Phys. Rev. {\bf D54} (1996) 1881; S.M. Bilenky, C.
Giunti, C. Kim, S. Petcov, Phys. Rev. {\bf D54}
(1996) 4444; J. Hellmig, H.V. Klapdor-Kleingrothaus, Z. Phys. {\bf A359}
(1997) 351; H. V. Klapdor-Kleingrothaus,
J. Hellmig, M. Hirsch, J. Phys. {\bf G24} (1998) 483; H. Minataka, O.
Yasuda, Phys. Rev. {\bf D56} (1997) 1692,
Nucl. Phys. {\bf  B523} (1998) 597; S.M. Bilenky, C. Giunti, C. W. Kim,
M. Monteno, Phys. Rev. {\bf D54} (1998) 6981,
hep-ph/9904328;
F.~Vissani,
%``Signal of neutrinoless double beta decay, neutrino spectrum and
%oscillation scenarios,''
JHEP {\bf 9906} (1999) 022;
T. Fukuyama, K. Matsuda,
H. Nishiura, Mod. Phys. Lett. {\bf A13} (1998) 2279; S.M. Bilenky, C.
Giunti, W. Grimus, hep-ph/9809368;
S. Bilenky, C. Giunti, hep-ph/9904328; S. Bilenky, C. Giunti, W. Grimus,
B. Kayser, S.T. Petcov,
Phys. Lett. {\bf B465} (1999) 193;
V. Barger, K. Whisnant, Phys. Lett. B{\bf 456} (1999) 194;
J.~R.~Ellis and S.~Lola,
%``Can neutrinos be degenerate in mass?,''
Phys. Lett. {\bf B458} (1999) 310;
G.C. Branco, M. N. Rebelo, J.I. Silva-Marcos, Phys. Rev. Lett. {\bf 82}
(1999) 683;
C. Giunti, Phys. Rev. {\bf D61} (2000) 036002;
R.~Adhikari and G.~Rajasekaran,
%``Constraints on mixing angles of Majorana neutrinos,''
Phys.\ Rev.\ D {\bf 61} (2000) 031301;
K.~Matsuda, N.~Takeda, T.~Fukuyama and H.~Nishiura,
%``CP violations in lepton number violation processes and neutrino
%oscillations,''
Phys.\ Rev.\ D {\bf 62} (2000) 093001; Y.~Farzan, O.~L.~Peres and A.~Y.~Smirnov,
%``Neutrino mass spectrum and future beta decay experiments,''
Nucl.\ Phys.\ B {\bf 612}, 59 (2001)
[arXiv:hep-ph/0105105].
%%CITATION = HEP-PH 0105105;%%

\bibitem{plb1} M.~Czakon, J.~Gluza and M.~Zralek,
Phys.\ Lett.\  {\bf B465} (1999) 211.

\bibitem{2n}
J.~N.~Bahcall, P.~I.~Krastev and A.~Y.~Smirnov,
%``Solar neutrinos: Global analysis and implications for SNO,''
JHEP {\bf 0105} (2001) 015;
J.~N.~Bahcall, M.~C.~Conzalez-Garcia and C.~Pena-Garay,
%``Global analysis of solar neutrino oscillations including SNO CC
%measurement,''
hep-ph/0106258;
M.~C.~Gonzalez-Garcia, P.~C.~de Holanda, C.~Pena-Garay and J.~W.~Valle,
%``Status of the MSW solutions of the solar neutrino problem,''
Nucl.\ Phys.\ B {\bf 573} (2000) 3;
A.~de Gouvea, A.~Friedland and H.~Murayama,
%``The dark side of the solar neutrino parameter space,''
Phys.\ Lett.\ B {\bf 490} (2000) 125.

\bibitem{3n} G.~L.~Fogli, E.~Lisi, A.~Marrone, D.~Montanino and
A.~Palazzo,
%``Atmospheric, solar, and CHOOZ neutrinos: A global three generation
%analysis,''
hep-ph/0104221;
G.~L.~Fogli, E.~Lisi, D.~Montanino and A.~Palazzo,
%``Three-flavor MSW solutions of the solar neutrino problem,''
Phys.\ Rev.\ D {\bf 62} (2000) 013002;
M.~C.~Gonzalez-Garcia, M.~Maltoni, C.~Pena-Garay and J.~W.~Valle,
%``Global three-neutrino oscillation analysis of neutrino data,''
Phys.\ Rev.\ D {\bf 63} (2001) 033005.

\bibitem{4n} C.~Giunti, M.~C.~Gonzalez-Garcia and C.~Pena-Garay,
%``Four-neutrino oscillation solutions of the solar neutrino problem,''
Phys.\ Rev.\ D {\bf 62} (2000) 013005;
V.~Barger, S.~Pakvasa, T.~J.~Weiler and K.~Whisnant,
%``Variations on four-neutrino oscillations,''
Phys.\ Rev.\ D {\bf 58} (1998) 093016


\bibitem{weinheimer}
J.~Bonn {\it et al.},
%``The Mainz neutrino mass experiment,''
Nucl.\ Phys.\ Proc.\ Suppl.\  {\bf 91} (2001) 273.
%%CITATION = NUPHZ,91,273;%%

\bibitem{lobashev} V.~M.~Lobashev {\it et al.},
%``Direct search for neutrino mass and anomaly in the tritium
%beta-spectrum: Status of 'Troitsk neutrino mass' experiment,''
Nucl.\ Phys.\ Proc.\ Suppl.\  {\bf 91} (2001) 280.

\bibitem{katr} for details see: http://www-ik1.fzk.de/tritium; 
A.~Osipowicz {\it et al.}  [KATRIN Collaboration],
%``KATRIN: A next generation tritium beta decay experiment with sub-eV  sensitivity for the electron neutrino mass,''
arXiv:hep-ex/0109033.
%%CITATION = HEP-EX 0109033;%%
\bibitem{half}  D.E. Groom et al., European Physical Journal C15 (2000)
1.
%``Review of particle physics. Particle Data Group,''
%H. V. Klapdor-Kleingrothaus et al, Annual Report Gran Sasso 2000
(2001);
%H. V. Klapdor- Kleingrothaus et al, MPI Heidelberg, Annual Report
1999-2000 (2001).

\bibitem{other}
W.~C.~Haxton and G.~J.~Stephenson,
%``Double Beta Decay,''
Prog.\ Part.\ Nucl.\ Phys.\  {\bf 12} (1984) 409;
M.~Doi, T.~Kotani and E.~Takasugi,
%``Double Beta Decay And Majorana Neutrino,''
Prog.\ Theor.\ Phys.\ Suppl.\  {\bf 83} (1985) 1;
M.~Hirsch, H.~V.~Klapdor-Kleingrothaus and O.~Panella,
%``Double beta decay in left-right symmetric models,''
Phys.\ Lett.\ B {\bf 374} (1996) 7;
H.~V.~Klapdor-Kleingrothaus, L.~Baudis, J.~Hellmig, M.~Hirsch, S.~Kolb,
H.~Pas and Y.~Ramachers,
%``Search for new physics with neutrinoless double beta decay,''
Nucl.\ Phys.\ Proc.\ Suppl.\  {\bf 70} (1999) 242;
H.~Pas, M.~Hirsch and H.~V.~Klapdor-Kleingrothaus,
%``Improved bounds on SUSY accompanied neutrinoless double beta decay,''
Phys.\ Lett.\ B {\bf 459} (1999) 450.
H.~Ejiri,
%``Double beta decays and neutrinos,''
Nucl.\ Phys.\ Proc.\ Suppl.\  {\bf 91} (2001) 255.
\bibitem{diff} V.~D.~Vergados,
%``Are Massive Majorana Neutrinos Cancelling Each Other In Neutrinoless
Double Beta Decay?,''
Phys.\ Rev.\ D {\bf 28} (1983) 2887;
C.~Greub and P.~Minkowski,
%``Heavy Majorana neutrinos in e- e- collisions,''
Int.\ J.\ Mod.\ Phys.\ A {\bf 13} (1998) 2363.

\bibitem{baudis}L.~Baudis {\it et al.},
%``Limits on the Majorana neutrino mass in the 0.1-eV range,''
Phys.\ Rev.\ Lett.\  {\bf 83}, 41 (1999).


\bibitem{nemo3} S.~Pirro {\it et al.},
%``Present status of MI-BETA cryogenic experiment and preliminary
%results  for CUORICINO,''
Nucl.\ Instrum.\ Meth.\ A {\bf 444}, 71 (2000).
\bibitem{ouore} E.~Fiorini,
%``CUORE: A cryogenic underground observatory for rare events,''
Phys.\ Rept.\  {\bf 307}, 309 (1998).

\bibitem{moon} H.~Ejiri, J.~Engel, R.~Hazama, P.~Krastev, N.~Kudomi and
R.~G.~Robertson,
%``Spectroscopy of double-beta and inverse-beta decays from Mo-100 for
%neutrinos,''
Phys.\ Rev.\ Lett.\  {\bf 85}, 2917 (2000).

\bibitem{exo} H.~V.~Klapdor-Kleingrothaus {\it et al.}  [GENIUS
Collaboration],
%``GENIUS: A supersensitive germanium detector system for rare events,''

hep-ph/9910205.



\bibitem{barger} V.~Barger, T.~J.~Weiler and K.~Whisnant,
%``Inferred 4.4-eV upper limits on the muon- and tau-neutrino masses,''
Phys.\ Lett.\ B {\bf 442} (1998) 255.

\bibitem{chooz} M.~Apollonio {\it et al.},
%``Limits on neutrino oscillations from the CHOOZ experiment,''
Phys.\ Lett.\ B {\bf 466} (1999) 415.

\bibitem{dm} M.~Czakon, M.~Zralek and J.~Gluza,
%``Are neutrinos Dirac or Majorana particles?,''
Acta Phys.\ Polon.\ B {\bf 30} (1999) 3121.


\bibitem{opera}  A.~G.~Cocco  [OPERA Collaboration],
%``The OPERA experiment at Gran Sasso,''
Nucl.\ Phys.\ Proc.\ Suppl.\  {\bf 85} (2000) 125;
K.~Lang  [MINOS Collaboration],
%``MINOS detectors for neutrino interactions,''
Nucl.\ Instrum.\ Meth.\ A {\bf 461} (2001) 290;
F.~Cavanna {\it et al.}  [ICARUS Collaboration],
%``ICANOE: Imaging and calorimetric neutrino oscillation experiment:
%Answers to questions and remarks concerning the ICANOE project,''
LNGS-P21-99-ADD-2.

\bibitem{era} V.~Barger,
%``Lepton flavor violating era of neutrino physics,''
hep-ph/0102052;
A.~Cervera, A.~Donini, M.~B.~Gavela, J.~J.~Gomez Cadenas, P.~Hernandez,
O.~Mena and S.~Rigolin,
%``Golden measurements at a neutrino factory,''
Nucl.\ Phys.\ B {\bf 579} (2000) 17
[Erratum-ibid.\ B {\bf 593} (2000) 731].

\bibitem{fact} V.~Barger, S.~Geer, R.~Raja and K.~Whisnant,
%``Long-baseline study of the leading neutrino oscillation at a
%neutrino  factory,''
Phys.\ Rev.\ D {\bf 62} (2000) 013004.
S.~Geer,
%``Neutrino factories: Physics,''
hep-ph/0008155;
B.~Richter,
%``Conventional beams or neutrino factories: The next generation of
%accelerator-based neutrino experiments,''
hep-ph/0008222;
%\bibitem{kam}
%M.~C.~Gonzalez-Garcia and C.~Pena-Garay,
%%``Global and unified analysis of solar neutrino data,''
%Nucl.\ Phys.\ Proc.\ Suppl.\  {\bf 91}, 80 (2000)
%[hep-ph/0009041].


\end{thebibliography}
\end{document}